\documentclass[modern]{aastex631}
\usepackage{makecell}

\newcommand{\newbf}[1]{#1}

\shorttitle{robostrategy}
\shortauthors{Blanton et al.}

\begin{document}

\title{{\tt robostrategy}: Field and Target Assignment Optimization\\
  in the Sloan Digital Sky Survey V}

\correspondingauthor{Michael R.~Blanton}
\email{mb144@nyu.edu}

\author[0000-0003-1641-6222]{Michael R.~Blanton}
\affiliation{Center for Cosmology and Particle Physics,
Department of Physics,
New York University, 
726~Broadway~Rm.~1005,
New York, NY 10003, USA}

\author[0000-0001-5926-4471]{Joleen K. Carlberg}
\affiliation{Space Telescope Science Institute, 3700 San Martin Dr., Baltimore, MD 21218}

\author[0000-0002-4459-9233]{Tom Dwelly}
\affiliation{Max-Planck-Institut f\"ur extraterrestrische Physik, Gie{\ss}enbachstra{\ss}e 1, 85748 Garching, Germany}

\author[0000-0003-3410-5794]{Ilija Medan}
\affiliation{Department of Physics and Astronomy, Vanderbilt University, Nashville, TN 37235, USA}

\author[0000-0003-0346-6722]{S. Drew Chojnowski}
\affiliation{Department of Physics, Montana State University, P.O. Box 173840, Bozeman, MT 59717-3840, USA}
\affiliation{NASA Ames Research Center, Moffett Field, CA 94035, USA}

\author[0000-0001-6914-7797]{Kevin Covey}
\affiliation{Department of Physics \& Astronomy, Western Washington University,
MS-9164, 516 High St., Bellingham, WA 98225, USA}

\author[0000-0001-9776-9227]{Megan C. Davis}
\affiliation{Department of Physics, University of Connecticut, 196A Auditorium Road Unit 3046, 
Storrs, CT 06269, USA}

\author[0009-0000-4049-5851]{John Donor}
\affiliation{Department of Physics and Astronomy, Texas Christian University, TCU Box 298840 
Fort Worth, TX 76129, USA }

\author[0000-0002-3956-2102]{Pramod Gupta}
\affiliation{Department of Astronomy, University of Washington, Box 351580,
Seattle, WA 98195, USA}

\author[0000-0002-4863-8842]{Alexander P. Ji}
\affiliation{Department of Astronomy and Astrophysics, University of Chicago, Chicago, IL 60637, USA}
\affiliation{Kavli Institute for Cosmological Physics, University of Chicago, Chicago, IL 60637, USA}
\affiliation{NSF-Simons AI Institute for the Sky (SkAI), 172 E. Chestnut St., Chicago, IL 60611, USA}

\author[0000-0001-7258-1834]{Jennifer A.~Johnson}
\affiliation{Department of Astronomy, The Ohio State University, Columbus, 140 W. 18th Avenue, OH 43210, USA} \affiliation{Center for Cosmology and Astroparticle Physics (CCAPP), The Ohio State University, 191 W. Woodruff Avenue, Columbus, OH 43210, USA}

\author[0000-0001-9852-1610]{Juna A.~Kollmeier}
\affiliation{Observatories of the Carnegie Institution for Science, 813 Santa Barbara Street, Pasadena, CA 91101, USA}

\author[0000-0003-2486-3858]{Jos{\'e} Sanchez-Gallego}
\affiliation{Department of Astronomy, University of Washington, Box 351580,
Seattle, WA 98195, USA}

\author[0000-0002-4454-1920]{Conor Sayres}
\affiliation{Department of Astronomy, University of Washington, Box 351580,
Seattle, WA 98195, USA}

\author[0000-0003-3769-8812]{Eleonora Zari}
\affiliation{Dipartimento di Fisica e Astronomia, Universit{\`a} di Firenze, Via G. Sansone 1, 50019, Sesto F.no (Firenze), Italy}
\affiliation{Max-Planck-Institut f\"ur Astronomie, Konigstuhl 17, D-69117,
Heidelberg, Germany}

\begin{abstract}
We present an algorithmic method for efficiently planning a long-term, large-scale
multi-object spectroscopy program.  The Sloan Digital Sky Survey V
(SDSS-V) Focal Plane System performs multi-object spectroscopy using
500 robotic positioners to place fibers feeding optical and infrared
spectrographs across a wide field. SDSS-V
uses this system to observe targets throughout the year at two
observatories in support of the science goals of its Milky Way Mapper
and Black Hole Mapper programs. These science goals require
observations of objects over time with preferred temporal spacings
(referred to as ``cadences''), which can
differ from object to object even in the same area of sky. {\tt
  robostrategy} is the software we use to construct our planned
observations so that they can best achieve the desired goals given the
time available as a function of sky brightness and local sidereal
time, and to assign fibers to targets during specific observations. We
use linear programming techniques to seek optimal allocations of time
under the constraints given. We present the methods and example
results obtained with this software.
\end{abstract}

\section{Introduction} \label{sec:intro}

Wide-field spectroscopic surveys play an irreplaceable role in better
understanding our universe \citep{york00a, colless01a, jones04a,
  eisenstein11a, blanton17a, kollmeier17a, abareshi22a, kollmeier25a}.  These
efforts are long term (years), expensive, and involving the time
and effort of dozens to hundreds of people. Therefore, we seek ways 
to conduct these surveys to maximize their scientific return.

When using a facility with a field of view much less than the full
available sky, astronomical surveys need to decide on a strategy
regarding how to allocate exposure time across the sky.  In the case
of targeted spectroscopic surveys, within each observation these
surveys also need to select specific targets. An optimal plan for such
a survey would maximize scientific return based on some metric, while
remaining within constraints regarding available observing time under
various conditions.

This paper addresses this problem in the context of the Sloan Digital
Sky Survey V (SDSS-V; \citealt{kollmeier17a, kollmeier25a}). SDSS-V
uses the 2.5-m Sloan Foundation Telescope (\citealt{gunn06a}) and the
100-inch du Pont Telescope (\citealt{bowen73a}), each outfitted with a
Focal Plane System with 500 robot positioners placing fibers in the
focal plane feeding an optical and an infrared spectrograph. With its
Milky Way Mapper (MWM) and Black Hole Mapper (BHM) programs, SDSS-V is
observing millions of targets, differentiated into ``cartons,'' each
associated with a specific target selection procedure. Different targets
have different cadence requirements regarding how to observe them
(number of observations, detailed timing, and which spectrograph). 
This system offers substantial observing flexibility (wide field, 
rapid time-domain response, optical and infrared). 
Because of the limited resource of observing time, we want to 
use this flexible system as efficiently as we can.

We describe here {\tt robostrategy}, which produces a plan for the
number of observations to make at what sky locations, under what sky 
brightness conditions, and at what hour angles, as well as specifying 
the targets for each observation. The plan is designed to fully satisfy 
the physical constraints of the system and the expected constraints on 
time availability.

\begin{figure}
\begin{center}
\includegraphics[width=0.85\textwidth]{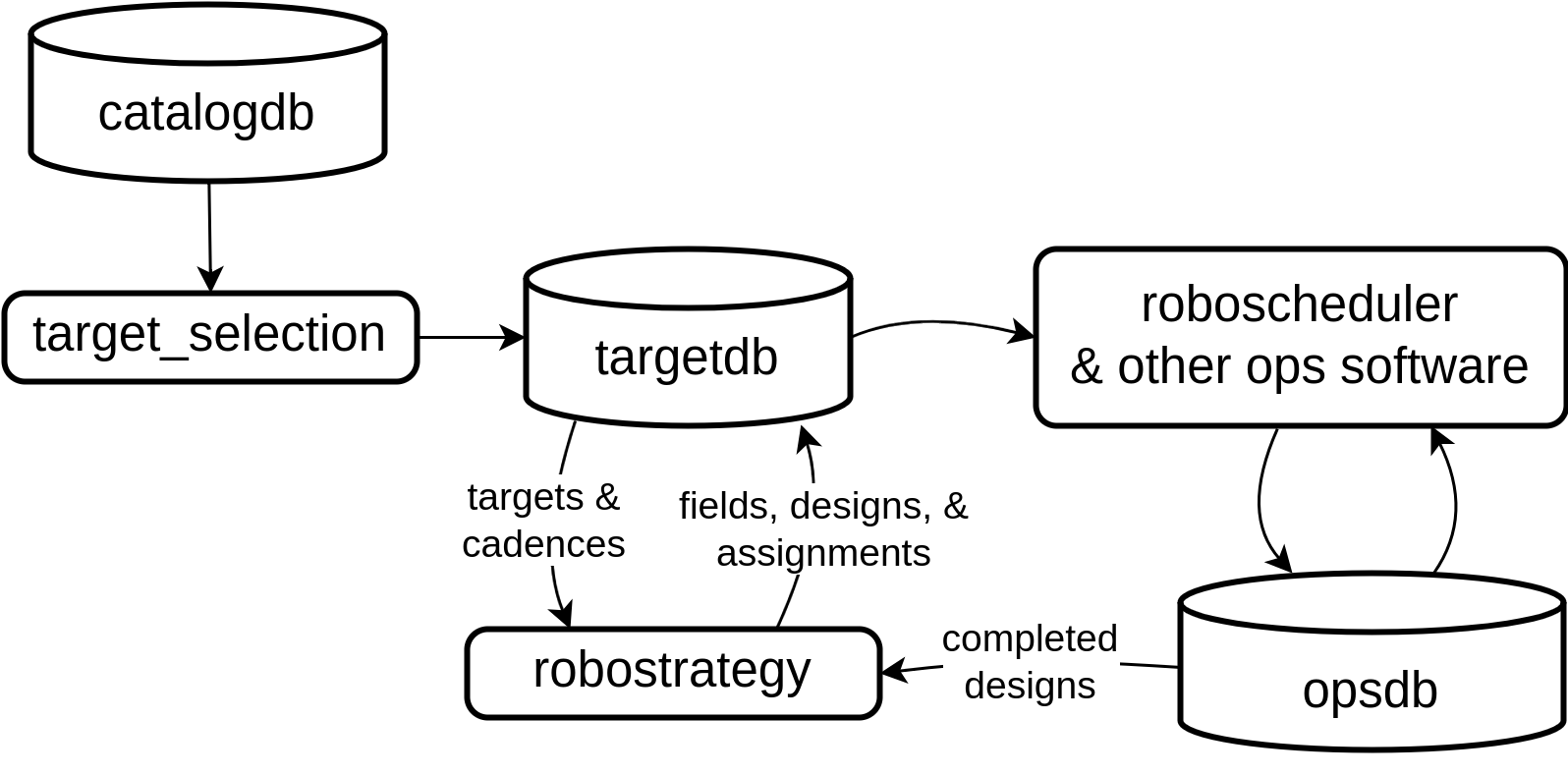}
\end{center}
\caption{\label{fig:high-level} Software context for {\tt
    robostrategy}. \newbf{Cylinders indicate databases, rectangles
    indicate processes and software, and arrows indicate a flow of
    information.} The fundamental catalog from which targets are drawn
  is in the database {\tt catalogdb}. The {\tt target\_selection}
  pipeline selects specific targets and specifies desired cadences,
  and stores the results in {\tt targetdb}. {\tt robostrategy} takes
  these targets as inputs, and delivers a set of field locations, a
  set of designs to observe for each field, a desired LST distribution
  in which to observe the designs in each field, and a set of
  assignments of robots to targets in each design. The operations
  software system, guided by the observers, uses this information from
  {\tt targetdb} to perform the observations and stores the
  as-observed set of configurations in {\tt opsdb}. When designs are
  deemed completed, this information is included in future runs of
  {\tt robostrategy}.}
\end{figure}

Figure \ref{fig:high-level} shows how {\tt robostrategy} fits into a
larger SDSS-V software context.  {\tt robostrategy} receives the
outputs of target selection in the form of targets and desired
cadences from a database {\tt targetdb}. Based on a high level
observing schedule and estimated exposure times and overheads, it
produces a set of fields and a number of designs in each field, a
desired Local Sidereal Time (LST) distribution for observing each
design, and assignments of fibers to targets for each design, which
are subsequently loaded back to {\tt targetdb}.  The {\tt
  roboscheduler} software, along with related operations software,
uses these designs and the LST plan as input to perform the real-time
scheduling at the telescope or in observing simulations.

\newbf{To perform the field allocation, we use linear programming, a
  well-established set of techniques which perform global maximization
  of an objective that depends linearly on its parameters, given
  linear constraints on those parameters. As we show below, the field
  allocation problem can be approximately expressed in such a form. A
  closely related type of problem, constraint programming (sometimes
  called integer programming), would exactly express the field
  allocation problem by restricting the solutions to integer values,
  but we found it too computationally expensive to implement.  To
  perform the fiber assignment, in most cases we use a greedy
  algorithm as explained below. However, in restricted cases we can
  efficiently express the problem as a constraint programming problem,
  (Section \ref{sec:constraintassignment}).}

%{\tt jaeger} loads the as-observed configurations of robots into {\tt opsdb}.

{\tt robostrategy} uses the robot positioner path software {\tt kaiju}
(\citealt{sayres21a}) to determine what combinations of robots can be
placed on what targets, and the design software {\tt mugatu} to
validate the designs and load them into the database
\citealt{medan25a}. All of these products depend on a coordinate
transformation package {\tt coordio} (\citealt{sayres22a}).

The problem that {\tt robostrategy} addresses is not unique to SDSS-V:
assigning fibers to targets using a specific multi-object fiber
system, and creating a plan to match observational resources. Among
other efforts, the Las Campanas Redshift Survey (LCRS;
\citealt{shectman96a}), the SDSS-I through -IV surveys
(\citealt{blanton03a}), the Two-degree Field Galaxy Redshift Survey
(2dFGRS; \citealt{colless01a}), and the Dark Energy Spectroscopic
Instrument (DESI; \citealt{abareshi22a}) have all addressed the fiber
assignment problem in various ways. This problem has also been
addressed in planning for the 4-metre Multi-Object Spectrograph
Telescope (4MOST; \citealt{tempel20a}). Our approach differs from the
methods described in many of those papers in its generality in dealing
with many potential target cadences observed in a given field.

Creating an allocation of time to fields has not often been approached
in as quantitative a fashion. For the SDSS-I through SDSS-IV surveys,
for 2dFGRS, and for the 4MOST planning published by \citet{tempel20b},
various \newbf{automated} methods were used to tune field locations and
coordinate observations between them to maximize
completeness. However, these \newbf{automated} methods proceeded without
regard to respecting the LST and lunation distribution, and the survey
strategies \newbf{(such as hour angle distributions of the
  observations)} were tuned to match observing resources more-or-less
by hand. By contrast, our work is aimed at using an optimization
approach to tune the number and cadence of observations per field,
while leaving the field positions fixed and not coordinating
observations between neighboring fields.

Section \ref{sec:nomenclature} describes our nomenclature.  Section
\ref{sec:fps} describes the constraints from hardware and astronomical
considerations defining how observations can occur and their cost in
observing time.  Section \ref{sec:inandout} describes the relevant
features of MWM and BHM that will affect our strategy, the inputs
provided to {\tt robostrategy}, and the desired outputs.  Section
\ref{sec:methodology} describes our methodology for defining the
survey strategy and assigning fibers.  Section
\ref{sec:implementation} describes the specific implementation choices
we used for this process at the beginning of SDSS-V.  Section
\ref{sec:summary} gives a brief summary.

\section{Nomenclature}
\label{sec:nomenclature}

We define the observational {\it plan} in terms of a set of {\it
  fields}. Each field is allocated a {\it field cadence}, and has
associated with it multiple {\it designs} that are created with a
particular {\it design mode}.  Each design is observed in one or more
{\it observations} under specific {\it observing mode} conditions.
The targets to observe are drawn from {\it cartons}, which assign to
each target its {\it category}, {\it instrument}, {\it target
  cadence}, {\it priority}, and {\it value}.

The definitions of these terms are as follows:
\begin{itemize}
\item Carton: A selection of targets of using a particular selection
  algorithm. Targets will be in one or more cartons.
\item Design: A definition of a planned observation, including 
  the fiber assignments.
\item Design mode: A specification of the desired design conditions,
  such as number and distribution of \newbf{flux calibration}
  standards and skies and conditions on brightnesses of science
  targets, standard targets, and brightness of \newbf{neighboring
    sources}.
\item Epoch: Within a target or field cadence, a set of observations
  that are expected to be observed back-to-back, under conditions and
  timing specified by the cadence.
\item Field: A center of the telescopic field of view along with a 
  position angle, for which there are one or more designs whose fiber
  assignments will be determined in coordination with each
  other.
\item Field allocation: A specification of which field cadences to
  use to observe each field. 
\item Field cadence: A specification of the desired observing mode,
  design mode, and timing of the epochs and observations within a
  field.
\item Instrument: One of the two spectrographs that objects may
  be assigned to (BOSS or APOGEE).
\item Observation: Within a target or field cadence, the basic
  observing unit, in SDSS-V corresponding to a 12 or 15 minutes of
  exposure time plus overhead, during which one BOSS spectrograph
  exposure and two APOGEE spectrograph exposures are taken
  simultaneously. Each observation uses a specific design.
\item Observing mode: A specification of the desired conditions
  (minimum distance from Moon, maximum predicted sky brightness level,
  minimum twilight angle, and maximum airmass).
\item Plan: Name referring to a specific run of robostrategy,
  encompassing a field allocation and associated fiber assignments.
\item Priority: An integer corresponding to the order of preference of
  targets during fiber assignments; lower numbers are assigned
  preferentially. 
\item Target cadence: A specification of the desired observing mode
  and timing of epochs and observations for a target.
\item Value: Relative scientific value of a target used in the context
  of allocating field cadences. 
\end{itemize}

In practice, the observing mode has a one-to-one relationship with the
design mode\newbf{, an implementation choice we made for simplicity}.

\newbf{The difference between priority and value is worth clarifying
  up front. The field allocation process performs a maximization over
  the sum of the values in order to allocate cadences to fields. There
  are some targets that we want to drive the field allocation, and the
  numbers and relative values of these targets end up determining how
  our fields are observed. When we perform assignments of targets to
  designs for a given field, the value plays no role. The targets are
  assigned in increasing order of their numeric priorities; for a
  given carton often the priorities are all equal, and in the greedy
  fiber assignment process are assigned in random order.}

\section{The SDSS-V Focal Plane System}
\label{sec:fps}

Two nearly identical SDSS-V Focal Plane Systems (FPS) are deployed on
two telescopes, the Sloan Foundation Telescope at Apache Point
Observatory (APO) and the du Pont Telescope at Las Campanas
Observatory (LCO). The two telescopes have different focal plane
shapes, and the Sloan Foundation Telescope is $f/5$ whereas the du
Pont Telescope is $f/7.5$.  Thus, the two telescopes have somewhat
different focal plane scales (measured on-axis, approximately 219.5 mm
deg$^{-1}$ at APO, and 328.9
%217.736
% 329.310
mm deg$^{-1}$ at LCO) and different usable fields of view (around 3
deg diameter at APO and around 2 deg diameter at LCO). They are both
telecentric.  The similarity in physical size and focal plane shape
means that nearly the same FPS system can be deployed on both to place
optical fibers precisely in the focal plane (\citealt{pogge20a}).

\begin{figure}
\begin{center}
\includegraphics[width=0.7\textwidth]{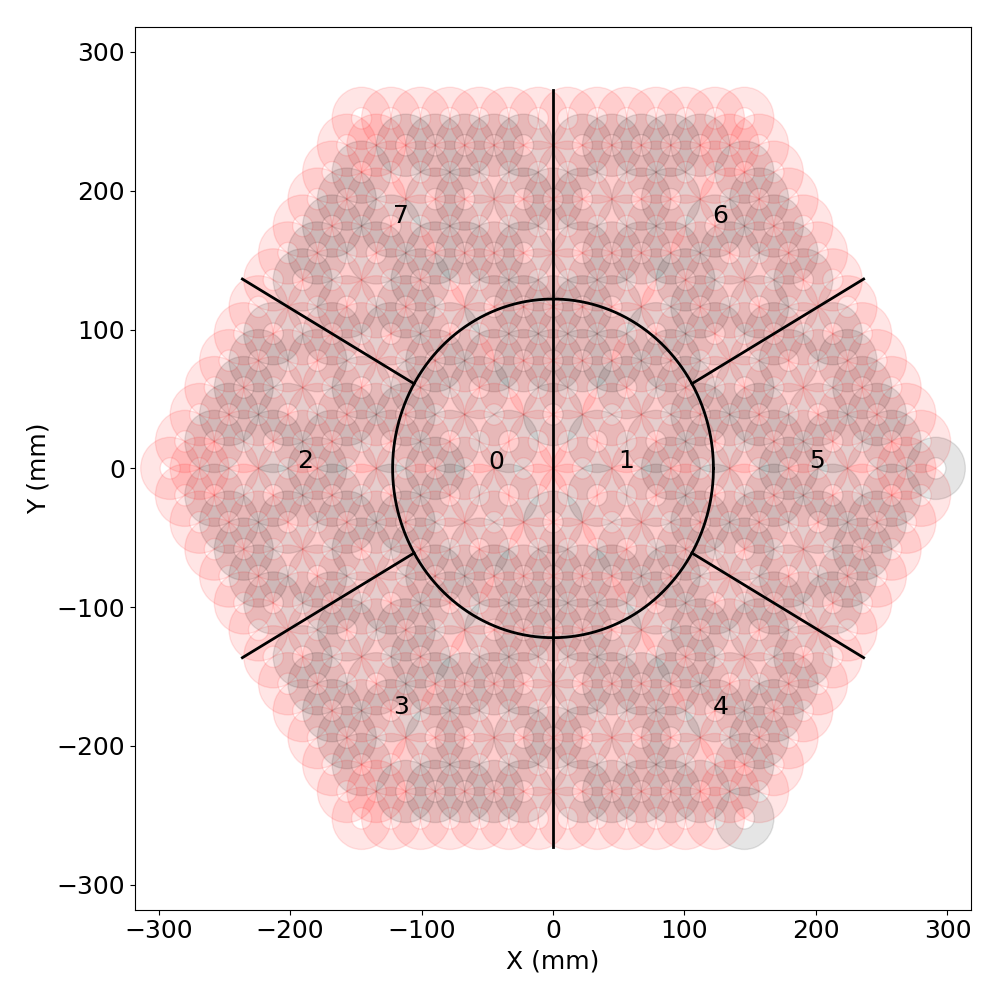}
\end{center}
\caption{\label{fig:fps} Layout of the focal plane for the FPS system
  used by SDSS-V. The layout shown is as-built for APO; the LCO layout
  is nearly identical.  The $X$ and $Y$ axes are position in the focal
  plane, with the boresight at $X=Y=0$ mm. We show each positioner as
  an annulus describing its patrol area. The pink annuli are the 298
  positioners that carry both BOSS and APOGEE fibers. The grey annuli
  are the 202 positioners that carry a BOSS fiber but not an APOGEE
  fiber.  The focal plane is divided into eight zones, as labeled, for
  the purposes of distributing APOGEE standard stars. The fiber reach 
  extends to around 315.5 mm in radius at the vertices of the hexagon, or 
  about 1.422 deg at APO and 0.953 deg at LCO, accounting for the 
  radial distortions at the edge of the field. This reach yields 
  a solid angle within
  the hexagon of 5.255 deg$^2$ at APO and 2.360 deg$^2$ at LCO. }
  % The two BOSS-only positioners in the outer ring correspond to the
  % Fabry-Perot Interferometer fibers for APOGEE.
\end{figure}

The FPS layout is shown in Figure \ref{fig:fps}. There are 500 robot
positioners, each of which carries three fibers
(\citealt{pogge20a}). One fiber feeds the optical BOSS spectrograph
(\citealt{smee13a}), one feeds the near infrared APOGEE spectrograph
(\citealt{wilson19a}), and the third can be backlit with a fiducial
light source. As illustrated in Figure \ref{fig:fps} only 298 of the
fibers actually feed the APOGEE spectrograph, due to the designed
capacity of its slit heads.

The robot positions have a two-arm design. Each arm can rotate
independently. The ``$\alpha$'' arm is further below the focal plane
and is half as long as the upper ``$\beta$'' arm. In this
configuration the robot can position the fibers anywhere in an
annulus. Figure \ref{fig:fps} shows the coverage of this system, which
nearly completely covers a hexagonal patch circumscribed in the usable
focal plane for both BOSS and APOGEE fibers. The $\beta$ arms are
about 3 mm in width and long enough to be able to collide with their
nearest neighbors, so there are constraints on the configurations of
the robot positioners that can be reached. The $\alpha$ arms cannot
collide with each other, and the fibers and other wires are arranged
so as not to interfere with each other.

\citet{sayres21a} describe the focal plane coverage and the collisions
in much greater detail. That work describes the {\tt kaiju} software
used to track the geometry of the system, predicting what
configurations are impossible because two $\beta$ arms would
collide. {\tt robostrategy} relies entirely on {\tt kaiju} to evaluate
these constraints.

{\tt kaiju} also determines the paths that the positioners need to
take to their final state to avoid a collision on the way to that
state. In rare cases, the final state has no collisions, but a
feasible path to it cannot be found. {\tt robostrategy} does not check
for this condition, which means that on rare occasions some fibers
will not be placed in practice.

\section{Inputs and Outputs to {\tt robostrategy}}
\label{sec:inandout}

Within SDSS-V, the FPS is used to perform the MWM and BHM multiobject
spectroscopy programs.  These programs are most efficiently performed
in a single coherent observing plan. Although BHM is primarily a dark time
program using the BOSS spectrograph, and MWM is primarily a bright
time program using the APOGEE spectrograph, for achieving their 
science goals both mappers to some extent do use both dark
time and bright time, as well as both spectrographs. Furthermore,
bright time targets can be observed in dark time. More generally, in
many fields neither mapper alone has sufficient appropriate targets for all
available fibers. These considerations drive us to a unified observing
plan for both mappers.
% NB, BHM does use bright time and the APOGEE spectrograph, in the
% bhm_csc program

\begin{figure}
\begin{center}
\includegraphics[width=0.99\textwidth]{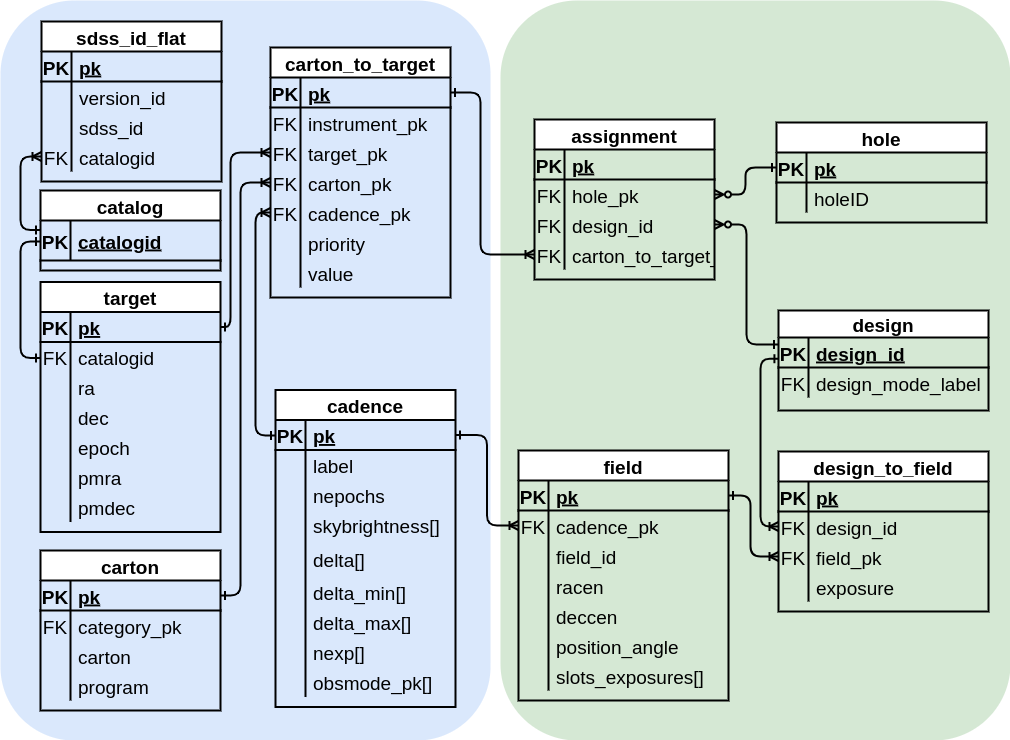}
\end{center}
\caption{\label{fig:db} Simplified schema of the database for SDSS-V
  planning. Primary keys are labeled ``PK'' and foreign keys used to
  join tables are labeled ``FK.''  Not all tables or table columns
  from the full schema are shown (which is why some foreign keys are
  not connected to tables).  The blue section represents the data
  produced prior to {\tt robostrategy}, and the green section is what
  {\tt robostrategy} determines. A detailed description is given in
  Section \ref{sec:inandout}.}
\end{figure}

The mapper teams select targets from a catalog stored in {\tt catalogdb},
which contains a unified table of objects drawn from numerous
photometric and spectroscopic surveys.  The target selection process
extracts the information from {\tt catalogdb} and stores the targeting
results in an associated {\tt targetdb} targeting database. 
SDSS-V target selection algorithms are described in more detail by
\citet{almeida23a} and \citet{sdss25a}. 
Figure \ref{fig:db} shows a simplified schema for the database
\newbf{including tables from both {\tt catalogdb} and {\tt targetdb}}; it omits a
number of tables and table columns.

The {\tt catalog} table is the result of cross-matching a large number
of parent catalogs. There are several versions of the cross-match that
we performed during the early phases of the survey. These
cross-matches linked Gaia (\citealt{prusti16a}), the Two Micron All
Sky Survey (2MASS; \citealt{skrutskie97a}), the SDSS imaging survey
(\citealt{york00a}), the Dark Energy Spectroscopic Instrument Legacy
Imaging Surveys (\citealt{dey19a}), and a number of other
catalogs. Each {\tt catalogid} corresponds to one or more entries in a
parent catalog.  All objects that can be targeted must have an entry
in the catalog table; this table is in {\tt catalogdb} and the primary
identifier within a cross-matching result is {\tt catalogid}.

To link objects across the cross-matches, we define a unique {\tt
  sdss\_id} for each unique object, and each {\tt sdss\_id} will
correspond to a different {\tt catalogid} for each cross-match. We
link different {\tt catalogid}s to the same {\tt sdss\_id} primarily
based on what original catalog entries they correspond to in the input
catalogs, occasionally relying on positional information. \newbf{The
  {\tt sdss\_id\_flat} table contains these identifications}.

\newbf{The} {\tt target\_selection} \newbf{pipeline} creates a set of
targets, at most one per {\tt catalogid}, which contain\newbf{s} relevant
information for targeting the object.  The target selection is
organized by carton, each of which corresponds to some set of
selection criteria. The science cartons are labeled as category {\tt
  science}, and the standard and sky calibration targets are labeled
as {\tt standard\_apogee}, {\tt standard\_boss}, {\tt sky\_apogee},
and {\tt sky\_boss}.

The {\tt carton\_to\_target} table is critical input into {\tt
  robostrategy}. It expresses which targets were selected by which
cartons. A target may be selected by any number of cartons. Each {\tt
  carton\_to\_target} entry specifies how the object is to be observed
for that carton: under what cadence, which instrument (BOSS or
APOGEE), the effective wavelength to use for placing the fiber, and
any offset in RA and Dec to apply. It also specifies the priority of
the target, which is used in assigning robots to targets, and the
value of the target, which is used in creating the field allocation.

The cadence definitions (used for both target and field cadences)
are also specified prior to running {\tt robostrategy}.  We describe 
the way that specification is defined in Section \ref{sec:cadence}.

The target cartons have different levels of importance, which will
define at what stage of {\tt robostrategy} fibers are assigned to
them. These levels of importance are:
\begin{itemize}
\item Science Requirements Document (SRD) Targets: Cartons around
  which the survey sucess is defined. The SRD cartons are further
  divided into Core and Shell; the Core target cartons are those that
  the field cadence allocation decisions are designed to optimize (that 
  is, they are assigned non-zero value in the algorithm).
\item Reassignment Targets: SRD target cartons which have the
  opportunity for extra observations or partial cadences, assigned
  after all SRD target assignment is complete.
\item Open Fiber Targets: Cartons used for open fibers after the SRD
  and reassignment stages are complete.
\item Filler Fiber Targets: Cartons used to fill any remaining spare
  fibers.
\end{itemize}
The filler fiber targets are in some sense just the lowest priority
open fiber targets, but they are handled in a separate stage because a
very large number of them are specified.  Within each level of
importance, the targets also are sorted according to a priority
number. 

At the end of target selection, the {\tt targetdb} contains in {\tt
  carton\_to\_target} the list of targets to be observed (including
repeats for targets in multiple cartons) and how they are meant to be
observed.

The task of {\tt robostrategy} is to define how these targets are
observed. The desired outputs are shown in the green region on the 
right hand side of Figure \ref{fig:db}. The survey plan is organized 
around the concept of a ``field,'' which corresponds to a particular 
field center and position angle.  The ideal desired distribution of 
LST and sky brightness associated with the field is stored in the 
array {\tt slots\_exposures}.

For each field, we may conduct multiple observations. For each
observation, we need a design expressing how each target is associated
with each fiber positioner.  There can be many designs per field, and
for technical reasons a design can be associated with multiple entries
in the field table (for example, multiple versions of the survey plan),
so there is a {\tt design\_to\_field} table expressing this
many-to-many relationship.

For each design, there is an associated set of assignments of
positioners to {\tt carton\_to\_target} entries. \newbf{Occasionally,
  different targets within the same carton have different priorities,
  cadences, or values associated with them, so the {\tt
    carton\_to\_target} table specifies these properties.} In detail
we assign each such target to the hole in the focal plane system that
holds the positioner (since the physical positioner itself may be
replaced during the survey). The relationship between designs,
targets, and holes is expressed in the {\tt assignment} table.

Thus, Figure \ref{fig:db} shows the inputs and outputs required for
{\tt robostrategy}, and the rest of this paper describes the procedure
of producing the latter from the former. \newbf{During development of
  plans, we do not actually load results into the database, which is
  reserved for results intended for operations.}

\section{{\tt robostrategy} Methodology}
\label{sec:methodology}

\subsection{Overview of Methodology}

The overall purpose of {\tt robostrategy} is to plan a set of cadences
for each field, and to assign fibers to targets within each
design. The assignment of cadences to fields (the field allocation) is
constrained by the estimated amount of available observing time
\newbf{over the during of the survey} as a
function of LST and lunation, and under those constraints the
procedure maximizes the total value of targets achieved in the field
allocation. The field allocation is not a moment-by-moment or
night-by-night plan, but instead just ensures that the estimated
amount of observing time resources is not exceeded.

There are two major pieces of {\tt robostrategy}, the field cadence
allocation and the fiber assignment. \newbf{ The final fiber
  assignment only occurs after the field cadence allocation, but the
  allocation does depend on a simplified version of the fiber
  assignments. We will explain the field cadence allocation first and
  the fiber assignment second.} To understand \newbf{either of} these
pieces, it is helpful to first understand {\tt robostrategy}'s
definition of a cadence.

\begin{figure}[!t]
\begin{center}
\includegraphics[width=0.9\textwidth]{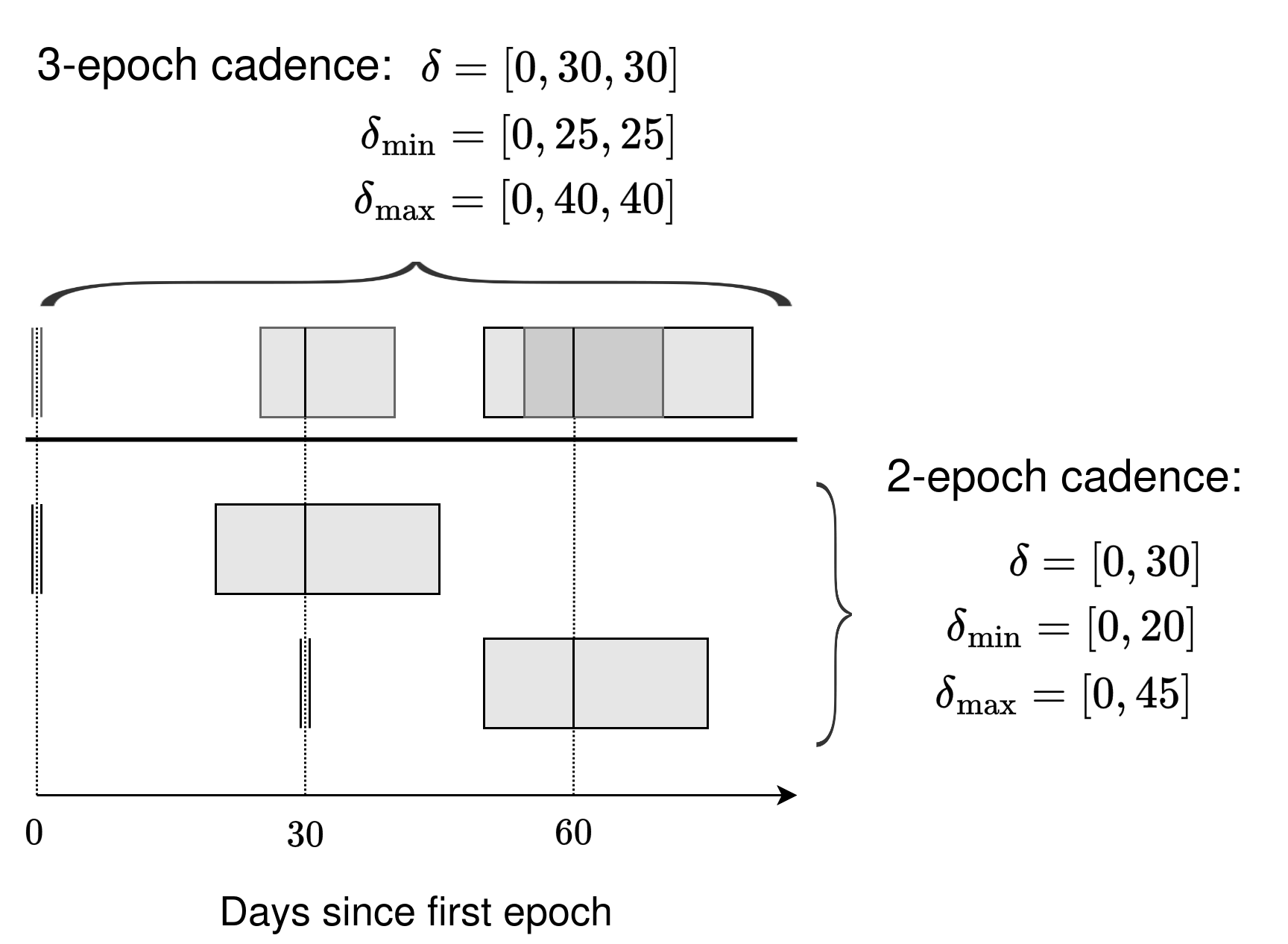}
\end{center}
\caption{\label{fig:cadence-timing} Example of the definition of two
different cadences, a 3-epoch cadence and a 2-epoch cadence, and the
ways a target with the 2-epoch cadence could be included in a field
with this 3-epoch cadence. After each epoch, the cadence
defines an acceptable timing of the next epoch, as a range of days
between $\delta_{\rm min}$ and $\delta_{\rm max}$; the ranges are
shown as the
horizontal bands for each cadence. The top row shows the 3-epoch
cadence; in the third epoch, we show both the range of acceptable
timing relative to the first epoch (lighter band) and relative to 
the second epoch, assuming it was observed at the preferred timing
$\delta$. Given a target with the 2-epoch cadence as shown, its 
timing requirements can be satisfied either by observing it in 
epochs 0 and 1, or epochs 1 and 2, of the 3-epoch field. But 
its timing requirements cannot be guaranteed by observing it in 
epochs 0 and 2.
For the 2-epoch cadence to ``fit'' into the 3-epoch cadence,
the requirements on the number of observations and the sky brightness
must also be satisfied.
In actual operations, the 
{\tt roboscheduler} will prefer timing close to $\delta$, but will not
necessarily strictly respect $\delta_{\rm min}$ and $\delta_{\rm max}$.}
\end{figure}

\subsection{Cadences}
\label{sec:cadence}

In the context of {\tt robostrategy}, a cadence is a set of
instructions regarding how to observe something, either a field or a
target.  The cadences are defined by the following properties (where
{\tt []} indicates an array of values, one for each epoch):
\begin{itemize}
\item {\tt nepochs}: number of distinct epochs;
\item {\tt nexp[]}: number of back-to-back observations planned in
  each epoch;
\item {\tt skybrightness[]}: maximum sky brightness (a number between 0 and 1) 
the epoch can be observed in;
\item {\tt delta[]}: desired delay between previous epoch and this one,
in days, or $-1$ if no particular timing is required;
\item {\tt delta\_min[]}: minimum delay between previous epoch and this one,
in days, or $-1$ if no particular timing is required;
\item {\tt delta\_max[]}: maximum delay between previous epoch and this one,
in days, or $-1$ if no particular timing is required.
\item {\tt obs\_mode\_pk[]}: name of observing mode and the design
  mode.
\end{itemize}

By ``observation,'' we mean a distinct set of assignments of robots to
targets to be observed in a nominal 15-minute or 12-minute period; in
this usage, each observation will correspond to a ``design\newbf{.''
  During
observations, each design is} instantiated as a ``configuration.''

The {\tt skybrightness} parameter is taken to be the fraction of the
Moon that is illuminated, or $0$ if the Moon is below the horizon, or
$1$ if it is between \newbf{the} ``bright time'' twilight
\newbf{limit} and \newbf{the} ``dark time'' twilight \newbf{limit (see
  Section \ref{sec:observing_time} for a description of these limits
  in the case of SDSS-V}. In practice we only distinguish between dark
time ($\le 0.35$) and bright time ($>0.35$). See Section
\ref{sec:observing_time} for more details on these definitions
specific to SDSS-V.

The observing mode defines the airmass limits, and also more detailed
observing condition requirements used during real-time scheduling.
The design mode defines bright magnitude limits for targets and
calibration fiber requirements (see Section \ref{sec:implementation}).
In practice, the design mode and observing mode names are the same.

Each target and field has a desired cadence. For example, using the
nomenclature ``number of epochs by number of observations per epoch,''
the {\tt dark\_2x4} cadence is defined as follows, in order to ensure
that the two epochs each have 4 observations, are taken in dark time,
and are at least about a year apart.

\begin{verbatim}
 nepochs = 2
 nexp = 4 4
 skybrightness = 0.35 0.35
 delta = 0.00 365.00
 delta_min = 0.00 300.00
 delta_max = 0.00 1800.00
 obsmode_pk = dark_monit dark_monit
\end{verbatim}

Figure \ref{fig:cadence-timing} demonstrates how a specific example of
a target cadence could ``fit'' inside a field cadence.  For any field
we observe, during the initial assignment phase, only targets with
cadences that fit into the field cadence can be observed. That means
that there must be a subset of epochs in the field cadence for which
the {\tt delta\_min} and {\tt delta\_max} requirements for the target
cadence epochs will be satisfied. In addition, there must be enough
observations in each epoch and the sky brightness criteria for the
targets must be satisfied. During later phases of assignment, targets
are given more flexibility to break cadence rules if needed to fill
the fibers efficiently.

\begin{figure}[!t]
\begin{center}
\includegraphics[width=0.9\textwidth]{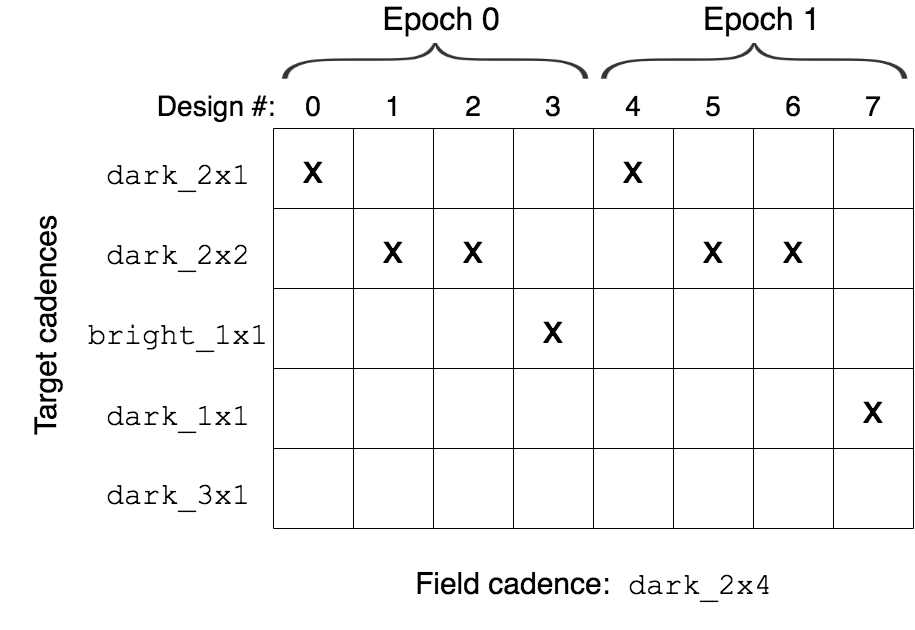}
\end{center}
\caption{\label{fig:cadences} Example of how targets with various
  cadences could be observed by a single robot in a field with a
  specific cadence. The nomenclature of the cadences is ``number of
  epochs by number of designs per epoch.''  The top four rows show
  cadences which can ``fit'' into the field epoch shown, i.e., that
  they have an equal or smaller number of epochs and an equal or
  smaller number of designs per epoch (assuming that the detailed
  timing of the epochs fits per Figure
  \ref{fig:cadence-timing}). Since in this case the field is a dark
  time field, the target with a bright time cadence has its sky
  brightness requirements fulfilled, whereas in a bright time field
  cadence, a target with a dark time cadence would not have its sky
  bright requirement fulfilled. The bottom row shows a target cadence
  that cannot be fulfilled in this field. }
\end{figure}

A single positioner in a field can be assigned to different targets in
different designs, as illustrated in Figure \ref{fig:cadences}, which
shows how 4 targets might be observed by a single positioner in a {\tt
  dark\_2x4} field cadence, if the individual target cadences each fit
into the field cadence.

The field cadences will be used in observations by {\tt
  roboscheduler}.  However, there are some differences in the way that
{\tt roboscheduler} interprets the cadences. First, {\tt
  roboscheduler} uses a more detailed description of the observing
conditions, using conditions on sky brightness based on the location
of the field, distance to the Moon, and Moon illumination
(\citealt{krisciunas91a}). The exact conditions are defined by {\tt
  obs\_mode\_pk}; many more details on these modes can be found in
\citet{medan25a}. Second, {\tt roboscheduler} uses the {\tt delta}
parameter to prefer a timing close to the specified delay, and also
does not strictly respect the {\tt delta\_max} values, in order to
improve observational flexibility.  \citet{donor24a} give a brief
description of how {\tt roboscheduler} works.

\subsection{Field cadence allocation}
\label{sec:allocation}

The field cadence allocation process determines how much time is spent
in which fields, and how that time should be spent in terms of a field
cadence.  It does not determine the exact planned timing of the
observations, which will depend on details of weather and other
contingencies. In our discussion here, we will assume that there is a
way of assigning targets to fibers that respects their relative
priorities, as described later in Section \ref{sec:assignment}.

The goal of the field cadence allocation is to decide how to allocate
time to fields in a way that maximizes the total scientific value
given the constraints on the amount of observing time (see Section
\ref{sec:observing_time}).

In {\tt robostrategy}, our procedure is as follows:
\begin{itemize}
\item Fix a number of field locations across the sky, with position
angles chosen to minimize the overlap between neighboring hexagonal 
fields.
\item For each field, indexed by $i$, perform target assignments using
  each potential cadence $j$ for that field, and calculate the total
  value $V_{ij}$ \newbf{over all targets} of each choice. \newbf{The
    values themselves are not used in the target assignment process,
    which depends only on priority.}
\item Under the constraints imposed by the available 
observing time, find the cadence choices that maximize 
the total value summed over all fields using a linear
programming technique (\citealt{matousek07a}).
\end{itemize}

For each field, we do not actually check all the defined cadences.
Instead, we only check the ones that are motivated by targets in the
field. For example, if there is a target in the field that has cadence
{\tt dark\_2x4}, that would motivate checking field cadences {\tt
  dark\_2x4} or {\tt dark\_10x4}, but not necessarily {\tt dark\_2x2},
since that field cadence cannot accommodate a {\tt dark\_2x4} target
cadence.

When we calculate the value of a particular set of assignments, not
all targets assigned necessarily count.  We have the freedom to assign
value to only certain cartons. However, we do need to account for all
targets (even those with no ``value'') in the assignment itself,
because even though targets may have no value for the purposes of
deciding upon field cadence allocation, they may have high priority
and affect the assignment of targets that do have value. We will
describe more about these choices for SDSS-V in Section
\ref{sec:targets}.

In order to express the constraints on observing time, we calculate
for the nominal program dates and for the observatory the available
bright and dark time in each hour of LST (see Section \ref{sec:observing_time}). 
We then assume an average
useful weather fraction. This leads to a 2$\times$24 array of slots
with the available number of hours per slot.

For each field and cadence choice, given the right ascension and
declination and the latitude of the observatory, we calculate which of
these LST and sky brightness bins it can be observed in, given the
maximum airmass limits and sky brightness values associated with the
cadence.  We also can calculate the efficiency of observing at each of
those LSTs.

We can then express the problem we want to solve as follows. We would
like to set the values of a variable $w_{ijk}$ equal to the number of
observations assigned to a given field, cadence for that field, and
observing slot, where:
\begin{itemize}
\item $i$ indexes the fields,
\item $j$ indexes the cadence choices of the field, and
\item $k$ indexes the LST and sky brightness slots.
\end{itemize}
If we determine $w_{ijk}$, we have determined how many observations we
will allocate to each cadence for each field, and what LSTs and sky
brightnesses those observations should be taken in. As we see below,
we will only allow one cadence to be assigned per field.

We will denote the following:
\begin{itemize}
\item $N_{ij}$: The number of designs required for cadence $j$ in
  field $i$.
\item $N_{{\rm b}, ij}$: The number of bright designs required for
  cadence $j$ in field $i$.
\item $N_{{\rm d}, ij}$: The number of dark designs required for
  cadence $j$ in field $i$.
\item $T_{k}$: The total amount of time available in each LST and sky
  brightness bin.
\item $V_{ij}$: As noted \newbf{in the procedure outline at the
  beginning of this section}, the value of fulfilling cadence $j$ in
  field $i$\newbf{, i.e. the sum of the value of the satisfied
    targets}.
\item $x_{ijk}$: Exposure time (same units as $T_k$) to apply for each
  design observed for this field, cadence, LST, and sky brightness
  value, to account for the cost of observing a field in a given
  way. This term accounts for the different overheads associated with
  single-observation and multi-observation epochs and the relative
  cost as a function of airmass.
\end{itemize}

The objective to maximize is the total value $V$ over all 
choices:
\begin{equation}
V = \sum_{ijk} \frac{w_{ijk}}{N_{ij}} V_{ij}.
\end{equation}
When $w_{ijk}/ N_{ij} =0$, the field is not observed at all
(no value), and when it $=1$, the field is fully observed
(full value). Because we are using a linear programming approach
and not a constraint programming approach,
in practice this ratio sometimes takes values in between 0 and 1;
we describe below how we handle these situations.

We need to impose constraints representing our observational
constraints. In what follows, we write down a set of constraints
appropriate to the approximate solution that we will seek with a
linear programming method (\citealt{matousek07a}); we would express
the constraint on assigning only one cadence field differently if we
were using an exact integer constraint programming method.

The easiest to understand constraints are:
\begin{eqnarray}
\label{eq:basic}
w_{ijk}&\ge& 0 \quad \mathrm{~nonnegative~numbers~of~observations;}\cr
%\sum_{k} w_{ijk} &\le& N_{ij} \quad \mathrm{~only~observe~as~much~as~necessary;}\cr
\sum_{jk} \frac{w_{ijk}}{N_{ij}} &\le& 1 \quad \mathrm{~only~one~cadence~per~field;}\cr
\sum_{ij} w_{ijk} x_{ijk} &\le& T_{k} \quad \mathrm{~respect~maximum~time~available~per~LST,~sky~brightness.}
\end{eqnarray}

In addition, we need the balance of dark and bright time 
to be correct, which is guaranteed with the following 
constraint:
\begin{equation}
\label{eq:mixed}
\frac{\sum_{k={\rm dark}} w_{ijk}}  
{\sum_{k={\rm bright}} w_{ijk}} >
\frac{N_{{\rm d}, ij}}{N_{{\rm b}, ij} }
\quad \mathrm{~have~at~least~enough~designs~in~dark~time.}
\end{equation}
In the context of linear or constraint programming, this condition is 
implemented as follows:
\begin{equation}
\label{eq:mixed_rewrite}
N_{{\rm b}, ij} \sum_{k={\rm dark}} w_{ijk}  -
N_{{\rm d}, ij} \sum_{k={\rm bright}} w_{ijk} \ge 0
\end{equation}

For some fields, we will guarantee that they are observed
(instead of assigned no cadence at all), a condition
which we impose with the constraint:
\begin{equation}
\label{eq:guarantee}
\sum_{jk} \frac{w_{ijk}}{N_{ij}} > (1-\epsilon) \quad \mathrm{~assign~some~cadence~for~field}~i,
\end{equation}
where we set \newbf{$\epsilon=10^{-3}$--$10^{-2}$} or similar.
Together with the second condition in Equation \ref{eq:basic}, the 
condition in Equation \ref{eq:guarantee} pins the field to have 
almost exactly one cadence. As we pointed out above, if we 
were using an integer contraint programming method, this guarantee
would be implemented differently.

The objective $V$ and all of the constraints above are expressible in
linear form.  A linear programming solution (details below) will find
the choice for all the $w_{ijk}$ that satisfies these constraints
(assuming they can be satisifed) and maximizes $V$. But the $w_{ijk}$
do not need to be integers, and multiple cadences can have non-zero
allocations of time in the same field. To arrive at a single cadence choice 
per field $i$, we first calculate the summed allocation of each cadence 
(across LST and sky brightness bins):
%its allocation to each cadence:
\begin{equation}
A_{ij} = \sum_k w_{ijk}
\end{equation}
and if any $A_{ij}>0$ we randomly choose the 
cadence $j$ to use according to the relative probabilities:
\begin{equation}
\label{eq:allocation}
p_{ij} = \frac{A_{ij}}{\sum_{j} A_{ij}}.
\end{equation}

At this stage, we have chosen a cadence $j$ for each field $i$ that
should satisfy the time constraints, accounting for the relative
cost of observing at different airmasses and with different conditions.
In principle the values of $w_{ijk}$ also tell us how to allocate
that time according to LST and sky brightness. However, 
there are a number of different choices for the allocation
across $k$ that have equivalent $V$, and that do not necessarily minimize
the total time utilized. In addition, the allocation across $k$ in
$w_{ijk}$ is affected by the fact that we have not solved the integer
problem.

Therefore we perform a second optimization, wherein for each
field $i$ we allow  $w_{ijk}>0$ only for one cadence $j$, corresponding
to that chosen in the first optimization. In the second optimization
we do not maximize $V$ (which is fixed by the cadence choices) but 
instead minimize the total observing time:
\begin{equation}
T = \sum_k T_k = \sum_{ijk} w_{ijk} x_{ijk}.
\end{equation}
For each field $i$, this second optimization tells us $w_{ijk}$, which
is how best to distribute designs across LST and sky brightness so
that the fields will efficiently fill the time. For example, if some
bright time designs need to be observed in dark time, this information
will be encoded in $w_{ijk}$. Similarly, the $w_{ijk}$ specify if some
designs need to be observed at high airmass (i.e. LSTs differing from
their RA).

This framework allows other constraints to be applied as well.  For
example, if we wanted to impose that some minimum number of targets
from some given carton be observed, that would be a linear constraint
that could be added; in fact the {\tt robostrategy} software includes 
this feature as a so-far unused option. Other
linear constraints are straightforward to impose.

In SDSS-V, there are $\mathcal{O}(10^4)$ fields, $\mathcal{O}(10^2)$
cadence possibilities per field, and $\mathcal{O}(50)$ (i.e. $24\times
2=48$) LST and sky brightness possibilities, leading to
$\mathcal{O}(5\times 10^7)$ variables. This large number of variables
would comprise a large constraint programming problem, which is why we
choose our approximate linear programming approach for each
optimization.

To solve the linear programming problems in these two optimizations we
use Google's OR-Tools Python interface, and the underlying Glop linear
programming
backend.\footnote{\url{https://developers.google.com/optimization}}

\subsection{Updating the Field Cadence Allocation}
\label{sec:updating_allocation}

During the course of a survey, target selection may undergo evolution
and estimates of overhead may change. In addition, the weather
outcomes and unplanned downtime can affect the observed LST
distribution, leading to progress that differs from that
planned. These events (all of which have happened in SDSS-V) mean that
we need a method to adjust our field cadence allocation to accommodate
them. This replanning process is generically a necessity for any
project, which may run less quickly, more quickly, or just differently
across the sky than originally planned.

We can use the same framework as above to update the field cadence
allocation, accounting for the existing observations. As we did
before, we perform target assignments for all relevant cadences for
all fields, using the techniques of Section
\ref{sec:updating_assignments} to account for previously observed
targets. For fields that have not yet received any observations, 
the procedure is identical to that described in Section \ref{sec:allocation}.

In updating the field cadence allocation, there is a new condition on
what cadences are relevant that we need to check. For each field, we
only consider a new cadence relevant if the already-observed designs
from the original cadence are consistent with being the first few
designs of the new cadence. We check whether these observations (as
planned for the original cadence) are consistent with the conditions
for the new cadence, and vice-versa. The basic ideas are, first, that
we should only check new cadences that have at least as many
observations as have already been taken, and, second, that for these
observations the new cadence should assume the same constraints as the
original cadence. It is the latter constraint that leads us to check
consistency in both directions; for example, if some observations were
required to be dark in the original cadence, the new cadence should
assume the same requirement. These additional constraints tend to 
reduce the pool of potential cadences that can be chosen for any field 
which has previously received some observations.

Once we have chosen the cadences to check, we need to adjust the
linear programming problem slightly. For each potential cadence some
of the dark and bright observations have already been fulfilled.
Accounting for these observations is slightly complicated due to the
fact that we are not treating this in a constraint programming
fashion. Each field $i$ will have already had part of cadence $j$
already completed. We only consider cadences that use all of the
previous observations, so the number can be written as $c_{i}$, and we
know how many bright and dark observations there were such that $c_{i}
= c_{{\rm d}, i} + c_{{\rm b}, i}$. Then the second constraint from
Equation \ref{eq:basic} will become:
\begin{equation}
\label{eq:basic_update}
\sum_{jk} \frac{w_{ijk}}{N_{ij} - c_i} \le 1
\end{equation}
This constraint guarantees that ``fulfilling the cadence'' will only
cost the additional, as yet unobserved, observations.  But it is
problematic when we are considering the original cadence and it has
already been fulfilled, because for that term the denominator is
zero. For this reason we use:
\begin{equation}
\label{eq:basic_update_rewrite}
\sum_{jk} \frac{w_{ijk}}{N_{ij} - c_i + \epsilon'} \le 1.
\end{equation}
This rewrite ensures that the terms corresponding to cadences that are
already satisfied are considered satisfied with a tiny allocation of
time.  For the fields guaranteed to have some cadence, we adjust
Equation \ref{eq:guarantee} similarly:
\begin{equation}
\label{eq:guarantee_update}
\sum_{jk} \frac{w_{ijk}}{N_{ij} - c_i + \epsilon'} > (1-\epsilon)
\end{equation}
We set \newbf{$\epsilon' \sim 10^{-4}$}, so that $\epsilon' < \epsilon$.

In addition, we again need the balance of dark and bright time to be
correct, and so Equation \ref{eq:mixed_rewrite} will become:
\begin{equation}
\label{eq:mixed_update}
\left[\left(N_{{\rm b}, ij} - c_{{\rm b}, i}\right)
\sum_{k={\rm dark}} w_{ijk}\right]
- \left[\left(N_{{\rm d}, ij} - c_{{\rm d}, i}\right)
\sum_{k={\rm bright}} w_{ijk}\right]
\ge 0
\end{equation}

Finally, in the selection of cadences from the linear programming
solution we still compute:
\begin{equation}
A_{ij} = \sum_k w_{ijk},
\end{equation}
which means that we select fields based on the fraction complete in
the {\it remaining} part of the field allocation.

The cost factor ({\tt xfactor} in the code) also needs to be adjusted
to account for the dark/bright balance remaining, since they have not
necessarily been completed at the same rate.

\subsection{Fiber assignment}
\label{sec:assignment}

\subsubsection{General considerations}

Given an assigned cadence for a field, we must assign fibers to
targets for each design of the cadence. This task must be performed
for each potential choice of cadence to estimate the value $V_{ij}$ of
each cadence, and then again for the final choice of cadence. In the
former case, we solve the assignment problem approximately in order to
keep the problem tractable. In this section we describe our method for
doing so.

For the final assignment, we need to respect the requirements on the
number of calibration fibers, meaning sky and standard targets for (in
most cases) the APOGEE and BOSS spectrographs.  For most calibration
categories, the requirement is just on the number of calibrators;
however for APOGEE standard stars the considerations are more
complicated as explained below.  In the assignments performed for the
field cadence allocation step above, we ignore the calibrations
entirely, to save computational effort.

\subsubsection{Greedy assignment basics: the usual case}
\label{sec:greedy}

There are five stages of assignment, described in more detail for SDSS-V in 
Section \ref{sec:assignment_stages}. The first four treat the different types of 
targets described in Section \ref{sec:inandout}: the SRD stage, the 
reassignment stage, the open fiber target stage, and the filler target stage. 
There is a final ``complete'' stage to fill any remaining unused fibers. 

Each of these stages typically uses the straightforward greedy
assignment algorithm we describe in this section.  In every case, we
sort the targets by priority, and in order of priority we seek to
assign each target to the design or designs that will satisfy the
target's cadence within the given field. Throughout these stages, the
software tracks how many calibration fibers can be assigned in order
to ensure the calibration requirements are respected; see the
description in Section \ref{sec:greedysrd}.

At a given priority level, we can have several different general types
of target cadences, which we treat separately and in the following
order:
\begin{itemize}
\item Targets with specific cadence requirements; i.e. that do not
  merely require some number of observations.
\item Targets that require one or more single bright time
  observations, but with no particular cadence.
\item Targets that require one or more dark time observations, but
  with no particular cadence.
\end{itemize}
The order matters, and effectively (within a given priority level)
favors the targets of types that appear earlier in the above list. Our
motivation for this choice is that targets with specific cadence
timing requirements are harder to satisfy, and the other targets can
more easily use the remaining available positions.

For targets with specific cadence requirements, we consider them one
by one. We first find the sets of epochs in the field cadence that
taken together can satisfy the target cadence.  Taking the earlier
sets of epochs first, we then ask if those epochs are available. To be
available, there must be enough designs for which a robot that can
reach the target is assignable. To be assignable, a robot must not be
assigned to any target already (unless it is with a spare calibration
fiber, see below), and the assignment must not cause a collision
(unless it is with a spare calibration fiber).  It can happen that for
a design a robot is already assigned to the same {\tt catalogid}, with
the same type of fiber, because the target appeared in a different
carton with a priority already considered; in such cases this
preexisting assignment is counted towards fulfilling the cadence
requirement.  For the first set of epochs for which there are enough
available robots, we assign them to the target in question.  For BOSS
targets, the software prefers to assign robots with only BOSS fibers,
since there are fewer robots with APOGEE fibers.

For targets that require designs with no particular cadence, the
process is effectively the same. However, we can implement it in a
more efficient fashion, because we do not have the complications of
checking the cadence timing or the number of observations per epoch.

By ``spare'' calibration target in the description above, we mean a
calibration target in excess of the requirements. During the process
described above, it occasionally happens that calibration targets are
assigned that exceed the requirements. If so, the software recognizes
this, so that if a robot is assigned to a particular category of
calibration target, and that calibration category has more than the
required number of targets, that robot is considered available.

When deciding whether a target is assignable to a robot in an
observation, in all cases we check if that assignment leads to either
the BOSS or APOGEE fiber on the robot being too close to a bright
neighbor. See Section \ref{sec:brightneighbor} for the exact
conditions that are checked.

\subsubsection{Implementation of greedy assignment in SRD stage}
\label{sec:greedysrd}

The first stage of assignment is of targets that are 
specified in the Science Requirements Document (the ``SRD'' stage).
In this stage the calibration targets are also assigned, and 
some care is taken to choose calibration targets that satisfy the
requirements, but in a manner that affects the SRD targets as little
as possible. The approach we describe here, though a bit complicated,
turns out to be a relatively simple implementation that does so; for 
example, assigning all of the calibration fibers first would limit 
the flexibility of the algorithm later, and assigning them all last would
make it often impossible to satisfy the calibration requirements.
The specific calibration requirements used for SDSS-V are
described in Section \ref{sec:calibration}. 

We start by first assigning the calibration targets before any science
targets are assigned. This step establishes the maximum number of
calibration targets possible that can be assigned in each category, the
``achievable'' number of calibration targets.  Sometimes this number
is smaller than the requirement; if so, {\tt robostrategy} will only
insist on reaching the achievable number, because it has no mechanism
to increase the number of available calibrators.  Some calibration
targets have a requirement that they are distributed across the focal
plane. To meet this requirement, we track the number of such targets
in each zone in Figure \ref{fig:fps} and define the required and
achievable numbers of targets in each zone separately (in the case of
SDSS-V, we require at least one {\tt standard\_apogee} target in each
zone, and other calibrations have no distribution requirements). After
the achievable number of calibrators is determined, we unassign all
the calibration targets.

Then for each priority level, we perform the assignment of
science targets (with details given below). After each priority
level, we then assign the calibration targets again; there will
be fewer calibration targets assignable after each priority level
as the science target occupy more and more of the fibers. 

After a given science priority level assignment, we may find that the
number of assigned calibration targets (and/or numbers in each zone)
for some calibration category and some design dips below the required
(or achievable) number. If so, we discard all of the science target
assignments at this priority level, and ``permanently'' assign the
given calibration category for the given design, i.e. so that future
science target assignments cannot ``steal'' those robots.  We then
reassign the science targets, try assigning the rest of the
calibration targets, and iterate until all designs have the required
(or achievable) number of calibration targets (with a maximum of four
iterations total per priority level).

The purpose of this procedure is to avoid assigning the calibration
targets all up front. Because of the robots' limited patrol area, if
the calibration targets were assigned up front, they would
unnecessarily limit our ability to observe science targets. For
example, with about 150 robots out of 500 allocated to calibration, if
they were all decided up front, the remaining 350 robots would not
cover the whole focal plane. The method described prioritizes
calibrations over science targets when necessary, but reduces the
degree to which science targets are excluded, since we can choose the best 150
robot to calibration target assignments that tend not to make the science targets
unreachable.

Our method for handling the calibration target requirements is 
appropriate for SDSS-V because the targets are
stratified into many different priority levels; if there were only one
or two priority levels, our method would have to be different.

\subsubsection{Implementation of greedy assignment in subsequent stages}
\label{sec:greedyother}

After the SRD stage, other observations of the SRD targets and other
targets are assigned to fill the remaining available fibers.  In these
subsequent stages, the greedy assignment is more straightforward
because we keep the calibration targets fixed.  We simply step through
priority levels and run the assignment described in Section
\ref{sec:greedy}.

\subsubsection{Updating Assignments Accounting for Previous Observations}
\label{sec:updating_assignments}

As the survey has proceeded, target selection has evolved. When the
list of targets and their priorities changes, we need to update the
{\tt robostrategy} results. In doing so, we need to account for
previous observations. This is slightly complicated for fields with
more than one design, since the unobserved designs should where
possible take advantage of the targets that already have observations.

In early updates to target lists and the assignment procedure, we had
only a small number of fields that had been started.  In those cases,
we only updated fields that had not been started, and did not update
any of the designs in the already-started fields. This procedure
avoided the complications mentioned above.  We used this approach in
the {\tt zeta-3} plan; see Section \ref{sec:history} and Table
\ref{table:history}. 

For later updates, we implemented a method to update future designs in
already-started fields. To do so, we first match the new set of
targets to the old targets and determine what successful observations
already exist. Then, for any design considered completed, in software
we assign the targets which have been determined to have been
successfully observed, and fix the assignments for those designs so
they cannot be changed (and lock any unassigned robot positioners in
those designs). Having done this, we can simply run the normal fiber
assignment process for the remaining designs. The software already has
to account for the possibility that a target was already assigned a
fiber under a different carton (see Sections \ref{sec:greedysrd} and
\ref{sec:greedy}); this feature allows it to take advantage of the
previously observed designs to fulfill cadences when it is making
assignments.

There are three subtleties that we account for. First, for those
targets with specific cadence requirements (those assigned in the
first bullet described in Section \ref{sec:greedy}), within each
priority level we first try to assign those targets which have already
been started. Second, for any targets with cadences that have been
started already, the code prefers to assign the original robot
positioner if possible; this preference tends to lead to an assignment
more consistent with the original assignment, reducing the number of
newly-created collisions that could disrupt the planned cadence for
other targets. Third, there are cases where a design has been observed
but for some reason an individual target has been deemed incomplete.
If other observations of the same target were successful, and the
failed observation is critical to completing a targets' cadence
(i.e. there is no way for the code to reassign it given the remaining
designs), then the code treats that observation as successful for that
target.

\subsubsection{Constraint programming assignment}
\label{sec:constraintassignment}

Here we describe a constraint programming-based assignment method that
we use in a limited number of cases.  The greedy assignment algorithm
has the virtue of computational efficiency, especially in the face of
multi-design fields with large numbers of targets. However, there is
no guarantee that it is optimal in terms of the number of
assignments. In certain cases it is both desirable and computationally
feasible to find the optimal solution. Here we describe the case where
we are performing assignments in only one design, so that the cadence
requirements are irrelevant.

The method proceeds in order of priority level. For any given priority
level, there will be science targets with previous assignments, there
will be science targets at the current level of priority, and there
will be calibration targets. As for the field cadence allocation step,
the method maximizes a linear objective while respecting linear
constraints.

We define a binary variable $q_{kl}$ for each pair of robot $k$ and
target $l$ for which the robot can reach the target.

We then define a linear objective $W$, which is to maximize the number of 
science targets at the current level of priority:
\begin{equation}
W = \sum_{kl} q_{kl}
\end{equation}
The constraints are as follows. First, each robot can only be assigned
to one target, and each target can only be assigned to one robot:
\begin{eqnarray}
\sum_{k} q_{kl} &\le& 1 
\quad \mathrm{~only~one~robot~per~target}\cr
\sum_{l} q_{kl} &\le& 1
\quad \mathrm{~only~one~target~per~robot}
\end{eqnarray}
Second, it may be that if robot $k$ is assigned to robot $l$, then
there will be a $\beta$ arm collision if robot $k'$ is assigned to
robot $l'$. We find all such cases and impose the condition
that that pair of assignments may not coexist:
\begin{equation}
q_{kl} + q_{k'l'} \le 1
\end{equation}
Third, all of the previously assigned science targets must be 
guaranteed a robot (any robot, not necessarily the robot they were 
previously assigned to):
\begin{equation}
\sum_k q_{kl} = 1 \quad \mathrm{~for~all~previously~assigned~science~targets~}l
\end{equation}
Fourth, there must be enough of each type of calibration 
target assigned:
\begin{equation}
\sum_{l{\rm ~in~}c} \sum_k q_{kl} \ge m_c \quad \mathrm{~for~each~calibration~target~category}~c
\end{equation}

This problem can be solved as an integer constraint programming
problem. With $\mathcal{O}(1000)$ targets and 500 robots, each of
which can reach about $\mathcal{O}(0.01)$ of the focal plane, this
leads to a few thousand variables and a similar number of constraints.
It must be solved for every priority level, of which there are tens.
This is a tractable problem with current tools.

Expanding this problem to multiple designs quickly increases the
number of variables, and in dense fields the number of targets can be
much higher. Including complicated cadences further complexifies the
problem enough that we have not implemented it for such cases.

To solve this integer constraint programming problems we use Google's
OR-Tools Python interface, and its CP-SAT
solver.\footnote{\url{https://developers.google.com/optimization}}

\section{Implementation in SDSS-V}
\label{sec:implementation}

\subsection{History of Implementation}
\label{sec:history}

Table \ref{table:history} lists the runs of {\tt robostrategy} used for
observations up to the writing of this paper. We plan future runs
throughout the survey as target selection evolves, as weather affects
the pattern of our progress over the sky, and as we better understand
or improve our efficiency.

The first set of SDSS-V observations were performed with the
spectroscopic plug-plate system that preceded the FPS, and were based
on targets from the {\tt 0.1.0} version of the cross-match of parent
catalogs. The plate observations did not use {\tt robostrategy} but
were planned in a fashion similar to previous SDSS programs. Planning
for the FPS observations does not explicitly account for the plate
program observations, except at the target selection stage. There was
also a pre-science {\tt robostrategy} run used for commissioning
observations, {\tt epsilon-7-core-0}, and some spectra in the SDSS
data releases are from these commissioning observations.

The first run of {\tt robostrategy} used in science operations was
{\tt zeta-0}, using tag {\tt 1.2.0}. This run was based on the {\tt
  0.5.0} cross-match of parent catalogs, and a suite of cartons
referred to as {\tt 0.5.2}. We performed the field cadence allocation
process and fiber assignment process, and FPS science observations at
APO began with this plan in February 2022.

The second run of {\tt robostrategy} was {\tt zeta-3}, using tag {\tt
  1.4.4}. This run was based on the {\tt 0.5.0} cross-match of parent
catalogs, and a suite of cartons referred to as {\tt 0.5.5}.  This run
used the same field cadence allocation determined in {\tt zeta-0}, and
only the fiber assignment was updated, and only on fields that had not
had any designs observed yet. Several minor improvements to the
efficiency of assigning targets were made, and some BHM cartons were
added to the reassignment stage. The major change implemented was to
the APOGEE standard target assignment, which imposed preferences for
color and brightness, and conditions on the distribution across the
field (as described below). FPS science observations at APO switched
to this plan in August 2022, and FPS commissioning observations at LCO
began with this plan in August 2022.

The third run of {\tt robostrategy} was {\tt zeta-4}, using tag {\tt
  1.4.4}. The only changes in this run were to the LCO field position
angles. They were restricted to the range 240$^\circ$ to 300$^\circ$,
since the as-built LCO system only allows the range 180$^\circ$ to
360$^\circ$; with the symmetry of the FPS, the 60$^\circ$ range is
sufficient. The fiber assignment was performed with this change; a
reassignment was necessary because the fiber robot collision
constraints do not have six-fold symmetry. FPS science observations at
LCO switched to this plan in September 2022.

The fourth run was {\tt eta-5}, the first run to include updates to
fields that had already been started (see Section
\ref{sec:updating_assignments}). This version incorporated a complete
revision of the targeting, using a new cross-match ({\tt 1.0.0}) and a
new definition of the cartons. We quickly identified an error in
targeting certain cartons of targets for probing radial velocity
variations, which led to a carton revision and an updated target list
for a fifth run, {\tt eta-6}. We recalibrated our model of the LCO
focal plane in early 2024, leading to a change in designs for {\tt
  eta-7}. We also added several cartons at this stage. After {\tt
  eta-7}, we realized several errors in the tracking of previously
observed targets, which {\tt eta-8} and {\tt eta-9} addressed.

The next run, {\tt theta-1}, was the first field cadence reallocation,
and used the methods described in Section
\ref{sec:updating_allocation}.  By the end of 2023, it had become clear that
the rate of progress for the survey was slower than the original plan
assumed. To account for this difference from expectations in the
reallocation, we took several steps. We increased the assumed
overhead. We also changed the planned coverage of the BHM All-Quasar 
Multi-Epoch Spectroscopy (AQMES) program, and reduced the allocation 
to BHM RM. The MWM-related allocations were reduced ``naturally'' because 
of the smaller number of observations available. However, we found that 
the most efficient allocation concentrated the observations to the Galactic 
Plane, because so many targets are available there, whereas stellar targets
are rare in the Galactic halo. Therefore, we altered the allocation
constraints to guarantee coverage of about 85\% of the sky, and we
reduced the number density of Galactic Plane targets. Finally, we made
two changes for the BHM SPIDERS program. We changed the cadence from
{\tt dark\_2x2} to {\tt dark\_1x3}, reducing the total exposure time
and eliminating the overhead associated with a second epoch. We also
imposed a declination-dependence to the value of observing BHM SPIDERS
targets at LCO, placing higher value to be near $\delta \sim
-20^\circ$ in the Northern Galactic Cap and near $\delta
\sim-45^\circ$ in the Southern Galactic Cap; this reweighting helped
{\tt robostrategy} choose a more contiguous footprint.

{\tt theta-3} implemented some changes to target lists, and also does
not require any explicit APOGEE sky fibers for any field with
$|b| > 20^\circ$; these fields all have enough fibers assigned to
relatively faint BOSS targets, for which the APOGEE
fiber is on a sufficiently dark part of the sky to be used as a sky
fiber.

The latest run is {\tt iota-1}, which used the target lists from {\tt
  theta-3} but accounted for a dramatic increase in survey speed,
especially at LCO, after improvements implemented around September
2024 in the operations software that places fibers.

There has been one special run, {\tt theta-2-boss-only-2}, which we
implemented for a several-week period at LCO \newbf{in January 2025},
while the APOGEE spectrograph was being serviced. This run only
allowed BOSS fibers and BOSS targets. We limited observations to
fields in three categories: first, assigned to dark time fields in
{\tt theta-1}; second, fields in LMC or SMC, to finish the BOSS
program on red giant branch stars there; and third, a special set of
cluster fields.

\begin{deluxetable}{lllll}
\tablecaption{\label{table:history} History up to March 2025 of
  queue-scheduled robostrategy runs}
\tablehead{
\colhead{Observatory} &
\colhead{Plan} &
\colhead{Tag} &
\colhead{Carton List} &
\colhead{Start Date} 
}
\startdata
APO & {\tt zeta-0} & {\tt 1.2.0} & {\tt 0.5.2} & February 2022 \cr
--- & {\tt zeta-3} & {\tt 1.4.4} & {\tt 0.5.5} & August 2022 \cr
--- & {\tt eta-5} & {\tt 1.5.10} & {\tt 1.0.2} & September 2023 \cr
--- & {\tt eta-6} & {\tt 1.5.10} & {\tt 1.0.3} & November 2023 \cr
--- & {\tt eta-8} & {\tt 1.5.10} & {\tt 1.0.4} & April 2024 \cr
--- & {\tt eta-9} & {\tt 1.5.14} & {\tt 1.0.4} & June 2024 \cr
--- & {\tt theta-1} & {\tt 1.6.5} & {\tt 1.0.5} & October 2024 \cr
--- & {\tt theta-3} & {\tt 1.6.11} & {\tt 1.0.6} & \newbf{February 2025} \cr
--- & {\tt iota-1} & {\tt 1.6.19} & {\tt 1.1.1} & \newbf{August 2025} \cr
LCO & {\tt zeta-4} & {\tt 1.4.4} & {\tt 0.5.5} & August 2022 \cr
--- & {\tt eta-5} & {\tt 1.5.10} & {\tt 1.0.2} & September 2023 \cr
--- & {\tt eta-6} & {\tt 1.5.10} & {\tt 1.0.3} & November 2023 \cr
--- & {\tt eta-7} & {\tt 1.5.10} & {\tt 1.0.4} & March 2024 \cr
--- & {\tt eta-8} & {\tt 1.5.11} & {\tt 1.0.4} & April 2024 \cr
--- & {\tt eta-9} & {\tt 1.5.14} & {\tt 1.0.4} & June 2024 \cr
--- & {\tt theta-1} & {\tt 1.6.5} & {\tt 1.0.5} & October 2024 \cr
--- & {\tt theta-2-boss-only-2} & {\tt 1.6.8} & {\tt 1.0.5-boss-only} & January 2025 \cr
--- & {\tt theta-3} & {\tt 1.6.11} & {\tt 1.0.6} & February 2025 \cr
--- & {\tt iota-1} & {\tt 1.6.19} & {\tt 1.1.1} & \newbf{August 2025} \cr
\enddata
\end{deluxetable}

\subsection{Targets}
\label{sec:targets}

\begin{figure}[!t]
\begin{center}
\includegraphics[width=0.99\textwidth]{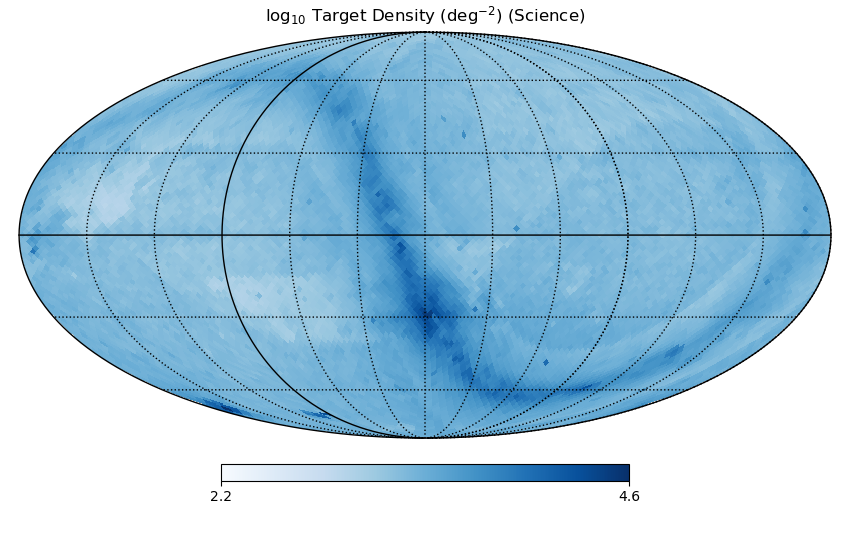}
\end{center}
\caption{\label{fig:targets} Science target distribution in Equatorial
  coordinates on a Molleweide projection.  The density is shown in
  HEALPix pixels (\citealt{gorski05a}) of approximately 1.8 deg on a
  side ({\tt NSIDE}=32 in HEALPix), and on a base-10 logarithmic scale
  in units of deg$^{-2}$.  RA$=270$ deg is at the center, RA=$0$ deg
  is indicated by the solid line, and RA increases to the left. The
  number counts are dominated by Galactic Genesis targets. This
  distribution corresponds to the targets used in the {\tt iota-1}
  plan.}
\end{figure}

Figure \ref{fig:targets} shows the distribution of targets on the sky
using a logarithmic stretch from {\tt iota-1}, showing the large
dynamic range in the target density across the sky.  The number
density is dominated by targets by the Galactic Genesis program. The
patterns of other targets on the sky are evident.

\begin{figure}[!t]
\begin{center}
\includegraphics[width=0.32\textwidth]{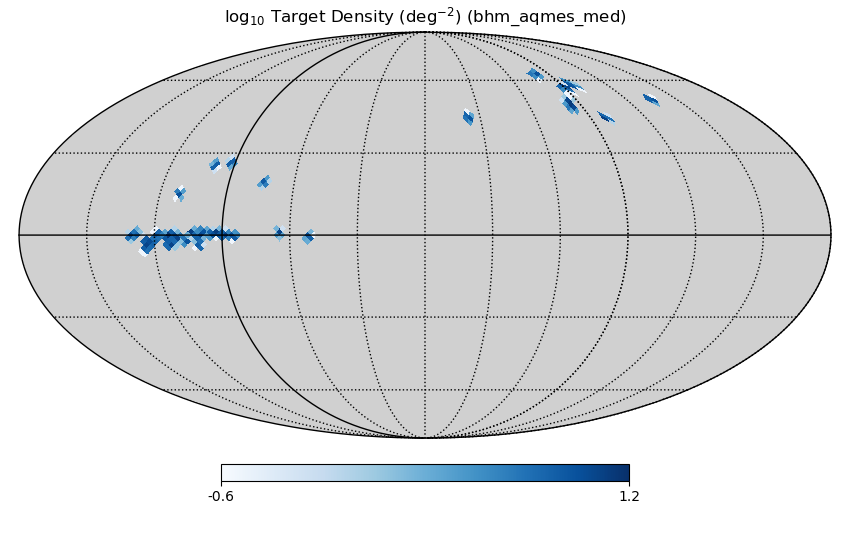}
\includegraphics[width=0.32\textwidth]{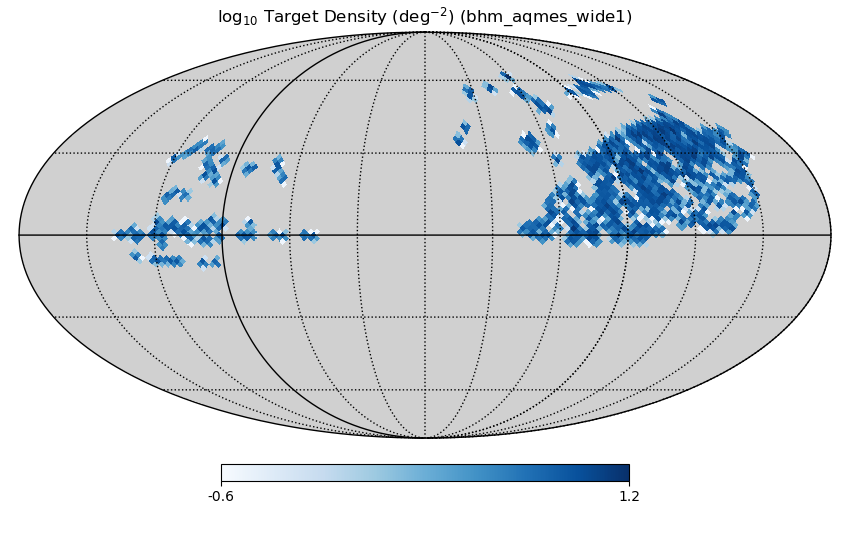}
\includegraphics[width=0.32\textwidth]{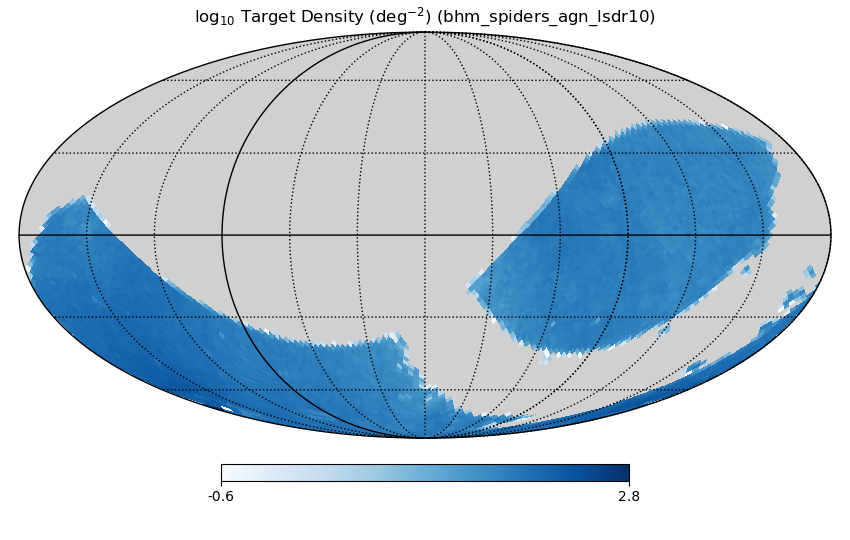}
\includegraphics[width=0.32\textwidth]{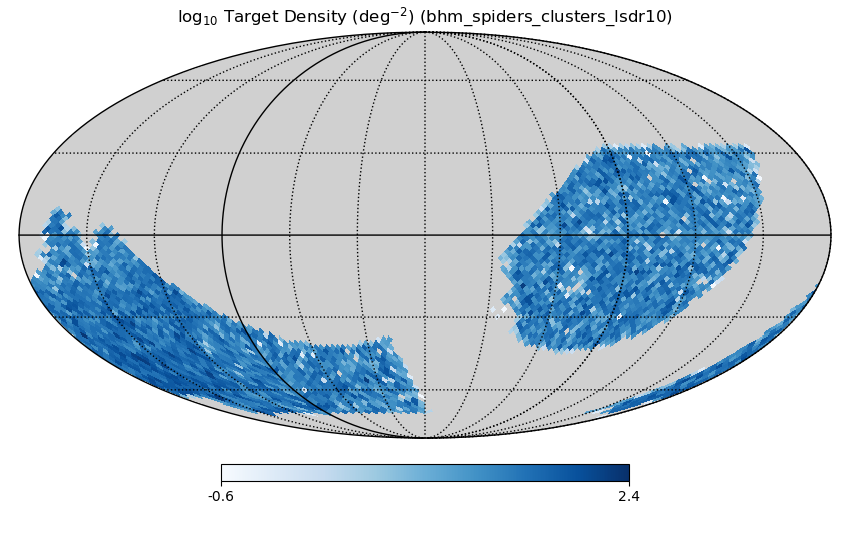}
\includegraphics[width=0.32\textwidth]{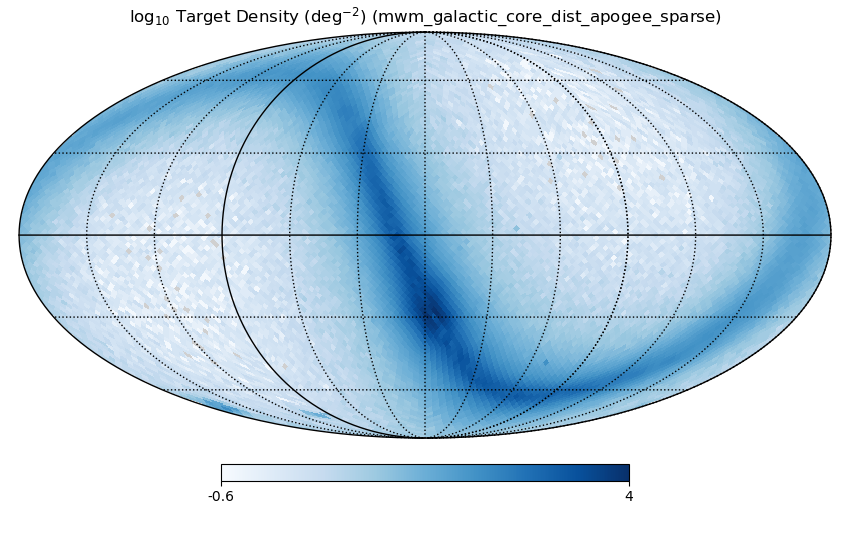}
\includegraphics[width=0.32\textwidth]{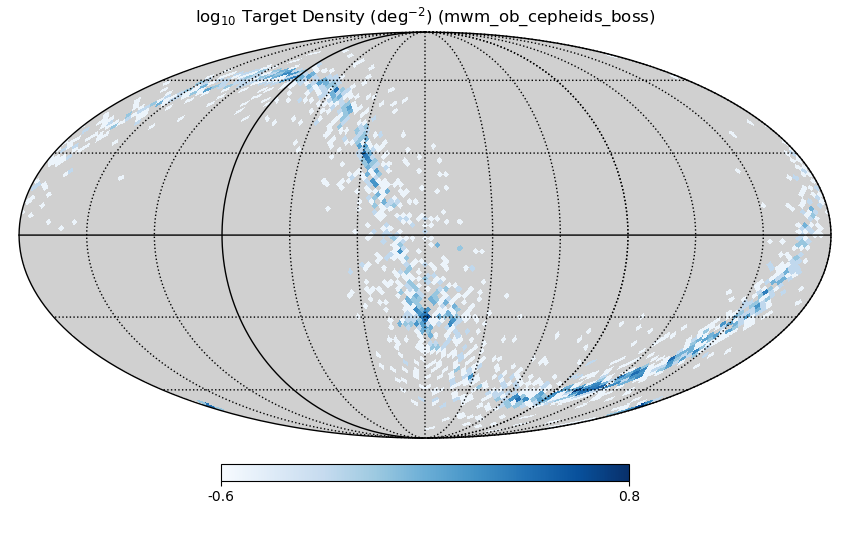}
\includegraphics[width=0.32\textwidth]{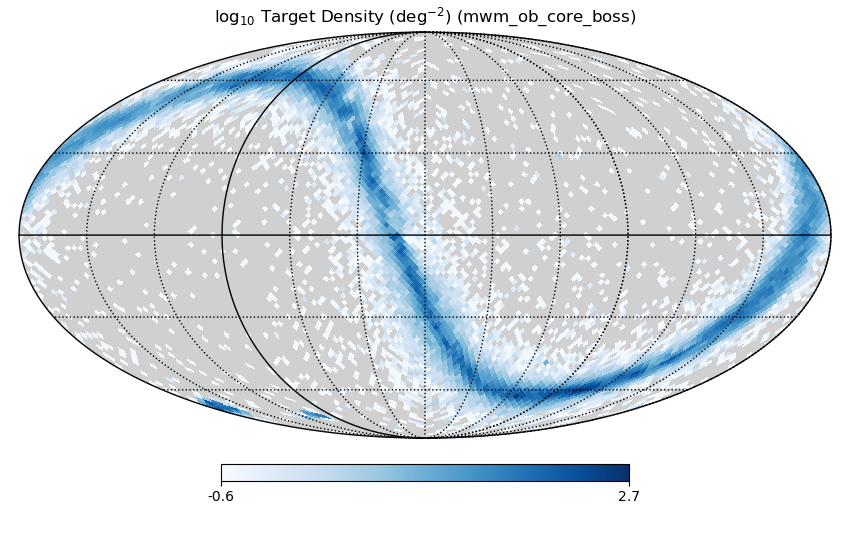}
\includegraphics[width=0.32\textwidth]{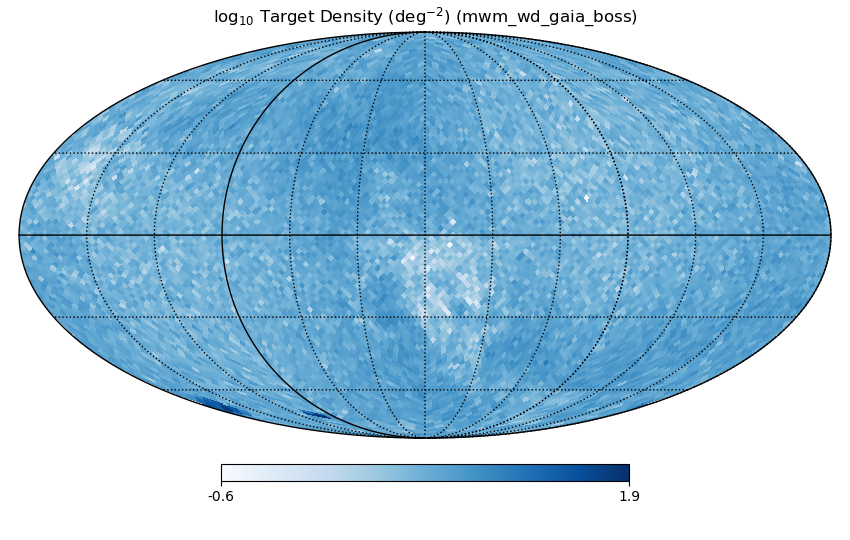}
\includegraphics[width=0.32\textwidth]{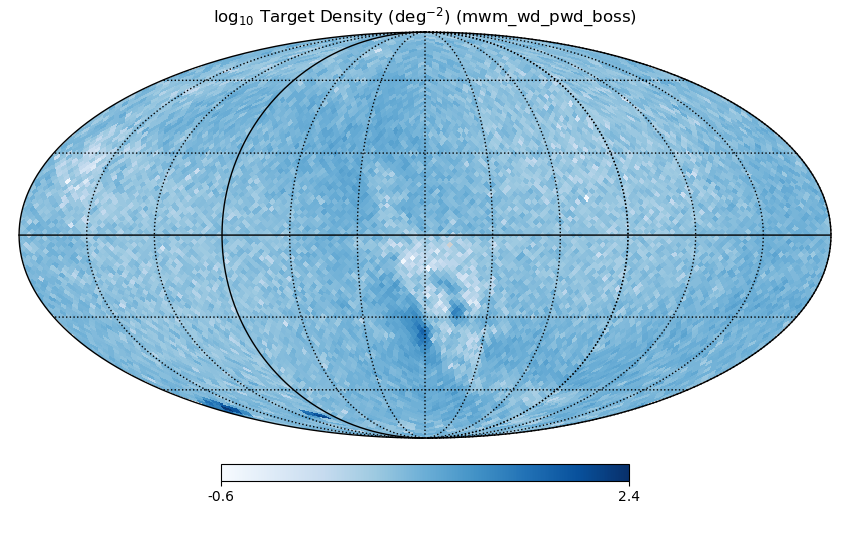}
\end{center}
\caption{\label{fig:corecartons} Core carton target distribution in
  Equatorial coordinates for the nine Core cartons for {\tt theta-1}
  listed in Table \ref{tab:corecartons_theta}, omitting the {\tt
    bhm\_rm\_core} carton. Each panel is similar to Figure
  \ref{fig:targets}. }
\end{figure}

Each target is given a value to be used in the field cadence
allocation process. For this purpose, only the Core SRD cartons are
given a non-zero value and therefore drive the allocation (the other
SRD cartons are referred to as Shell). Table
\ref{tab:corecartons_zeta} lists the Core cartons used in the time
allocation associated with the {\tt zeta-0} plan\newbf{; we will not
  list the much more numerous Shell cartons here}. In the
reallocations for {\tt theta-1} \newbf{and {\tt iota-1}}, we used a
slightly different set of choices of core cartons and value weights;
Table\newbf{s \ref{tab:corecartons_theta} and
  \ref{tab:corecartons_iota} list} these choices. Figure
\ref{fig:corecartons} shows the individual distribution of each of
these cartons across the sky, for the {\tt theta-1} case. \newbf{The
  chosen values are the primary way in which the user controls the
  behavior of {\tt robostrategy}'s field allocation, and we perform a
  high degree of tuning and testing of these values to achieve the
  desired results}.

Each target is also given a priority that is used when targets are
assigned to fibers for a given cadence in a given field. Lower
priority numbers are considered first in the assignment process
(i.e. priority 1000 will be given preference over priority 2000).

The Core cartons do not always have the highest priority. For example,
there are rare targets which we need to give high priority to in order
to get a large enough sample (they are favored when it comes to fiber 
assignment), but which we do not want to drive the
allocation of observing time across fields.

In addition to the science targets, there are also calibration targets
in four categories: 
{\tt standard\_apogee},
{\tt standard\_boss},
{\tt sky\_apogee}, and
{\tt sky\_boss}.

\begin{deluxetable}{lllll}
\rotate
\tablecaption{\label{tab:corecartons_zeta} Core target cartons for
  the SDSS-V FPS program for first field cadence allocation ({\tt
    zeta-0})}
\tablehead{
\colhead{Name ({\tt v0.5})} &
\colhead{Name ({\tt v1})} &
\colhead{Description} &
\colhead{Cadences ({\tt v1})} &
\colhead{Value}
}
\startdata
{\tt bhm\_aqmes\_med} & {\tt bhm\_aqmes\_med} & AQMES Medium & {\tt dark\_10x4} & 40 \cr
{\tt bhm\_aqmes\_wide2} & {\tt bhm\_aqmes\_wide2} & AQMES Wide & {\tt dark\_2x4} & 40 \cr
{\tt bhm\_rm\_core} & {\tt bhm\_rm\_core} & Reverberation Mapping & \makecell{{\tt dark\_178x8}\\  {\tt dark\_100x8}} & 40 \cr
{\tt bhm\_spiders\_agn\_lsdr8} & {\tt bhm\_spiders\_agn\_lsdr10} & X-ray AGN & 
 \makecell{{\tt dark\_flexible\_2x2} \\ {\tt dark\_flexible\_2x1} \\ {\tt bright\_flexible\_2x1}} & 20 \cr
{\tt bhm\_spiders\_agn\_ps1dr2} & {\tt bhm\_spiders\_agn\_gaiadr3} & X-ray AGN &  
 \makecell{{\tt dark\_flexible\_2x2} \\ {\tt dark\_flexible\_2x1} \\ {\tt bright\_flexible\_2x1}} & 20 \cr
{\tt mwm\_galactic\_core} & {\tt mwm\_galactic\_core\_dist\_apogee} & Galactic Genesis & 
 {\tt bright\_1x1} & 5 \cr
{\tt mwm\_ob\_core} & \makecell{{\tt mwm\_ob\_core\_boss}\\ {\tt mwm\_ob\_cepheids\_boss}} & OB Stars & 
 {\tt bright\_3x1} & 10 \cr
{\tt mwm\_tess\_ob} & {\tt manual\_mwm\_tess\_ob\_apogee} & TESS CVZ OB stars & 
 \makecell{{\tt bright\_8x1}\\ {\tt bright\_8x2}\\ {\tt bright\_8x4}} & 100 \cr
{\tt mwm\_wd\_core} & \makecell{{\tt mwm\_wd\_pwd\_boss}\\ {\tt mwm\_wd\_gaia\_boss}} & White Dwarfs 
  & {\tt dark\_2x1} & 10 
\enddata
\end{deluxetable}

\begin{deluxetable}{llll}
\rotate
\tablecaption{\label{tab:corecartons_theta} Core targets cartons for
  the SDSS-V FPS program, for \newbf{the} second field cadence allocation ({\tt
    theta-1})}
\tablehead{
\colhead{Name ({\tt v1})} &
\colhead{Description} &
\colhead{Cadences ({\tt v1})} &
\colhead{Value}
}
\startdata
{\tt bhm\_aqmes\_med} & AQMES Medium & {\tt dark\_10x4} & 40 \cr
{\tt bhm\_aqmes\_wide2} & AQMES Wide & {\tt dark\_2x4} & 40 \cr
{\tt bhm\_aqmes\_wide1} & AQMES Wide & {\tt dark\_1x4} & 30 \cr
{\tt bhm\_rm\_core} & Reverberation Mapping & \makecell{{\tt dark\_178x8}\\  {\tt dark\_100x8}} & 40 \cr
{\tt bhm\_spiders\_agn\_lsdr10} & X-ray AGN & 
 \makecell{{\tt dark\_flexible\_2x2} \\ {\tt dark\_flexible\_2x1}
   \\ {\tt bright\_flexible\_2x1}} & 200 \cr
{\tt bhm\_spiders\_agn\_lsdr10\_d3} & X-ray AGN & 
 {\tt dark\_flexible\_3x1} & 2000 \cr
{\tt bhm\_spiders\_clusters\_lsdr10} & X-ray \newbf{Cluster Galaxies} & 
 \makecell{{\tt dark\_flexible\_2x2} \\ {\tt dark\_flexible\_2x1}
   \\ {\tt bright\_flexible\_2x1}} & 200\tablenotemark{a} \cr
{\tt bhm\_spiders\_clusters\_lsdr10\_d3} & X-ray \newbf{Cluster
  Galaxies} & 
 {\tt dark\_flexible\_3x1} & 2000\tablenotemark{a} \cr
{\tt mwm\_galactic\_core\_dist\_apogee\_sparse} & Galactic Genesis & 
 {\tt bright\_1x1} & 5 \cr
\makecell{{\tt mwm\_ob\_core\_boss}\\ {\tt mwm\_ob\_cepheids\_boss}} & OB Stars & 
 {\tt bright\_3x1} & 20 \cr
\makecell{{\tt mwm\_wd\_pwd\_boss}\\ {\tt mwm\_wd\_gaia\_boss}} & White Dwarfs 
  & {\tt dark\_2x2} & 20  \cr
\makecell{{\tt mwm\_wd\_pwd\_boss\_1x3}\\ {\tt mwm\_wd\_gaia\_boss\_1x3}} & White Dwarfs 
  & {\tt dark\_1x3} & 20 
\enddata
\tablenotetext{a}{A declination-dependent factor is applied to prefer 
targets near $\delta \sim
-20^\circ$ in the Northern Galactic Cap and near $\delta
\sim-45^\circ$ in the Southern Galactic Cap}
\end{deluxetable}

\newbf{
\begin{deluxetable}{llll}
\rotate
\tablecaption{\label{tab:corecartons_iota} Core targets cartons for
  the SDSS-V FPS program, for the third field cadence allocation ({\tt
    iota-1})}
\tablehead{
\colhead{Name ({\tt v1})} &
\colhead{Description} &
\colhead{Cadences ({\tt v1})} &
\colhead{Value}
}
\startdata
{\tt bhm\_aqmes\_med} & AQMES Medium & {\tt dark\_10x4} & 40 \cr
{\tt bhm\_aqmes\_wide2} & AQMES Wide & {\tt dark\_2x4} & 40 \cr
{\tt bhm\_aqmes\_wide1} & AQMES Wide & {\tt dark\_1x4} & 30 \cr
{\tt bhm\_rm\_core} & Reverberation Mapping & {\tt dark\_178x8} & 40 \cr
{\tt bhm\_spiders\_agn\_lsdr10} & X-ray AGN & 
 \makecell{{\tt dark\_flexible\_2x2} \\ {\tt dark\_flexible\_2x1}
   \\ {\tt bright\_flexible\_2x1}} & 10 \cr
{\tt bhm\_spiders\_agn\_lsdr10\_d3} & X-ray AGN & 
 {\tt dark\_flexible\_3x1} & 10 \cr
{\tt bhm\_spiders\_clusters\_lsdr10} & X-ray \newbf{Cluster
  Galaxies} & 
 \makecell{{\tt dark\_flexible\_2x2} \\ {\tt dark\_flexible\_2x1}
   \\ {\tt bright\_flexible\_2x1}} & 10\tablenotemark{a} \cr
{\tt bhm\_spiders\_clusters\_lsdr10\_d3} & X-ray \newbf{Cluster
  Galaxies} & 
 {\tt dark\_flexible\_3x1} & 10\tablenotemark{a} \cr
{\tt mwm\_galactic\_core\_dist\_apogee\_sparse} & Galactic Genesis & 
 {\tt bright\_1x1} & 5 \cr
{\tt mwm\_wd\_pwd\_boss\_2x1} & White Dwarfs 
  & {\tt dark\_2x1} & 4  \cr
{\tt mwm\_wd\_pwd\_boss\_2x1} & White Dwarfs 
  & {\tt dark\_2x1} & 4 
\enddata
\tablenotetext{a}{A declination-dependent factor is applied to prefer 
targets near $\delta \sim
-20^\circ$ in the Northern Galactic Cap and near $\delta
\sim-45^\circ$ in the Southern Galactic Cap. Relative to the factor
applied in {\tt
  theta-1} (Table \ref{tab:corecartons_iota}), this declination dependent factor has a much stronger
declination dependence and reaches much higher values.}
\end{deluxetable}
}

\subsection{Field positions and position angles}

\begin{figure}[!t]
\begin{center}
\includegraphics[width=0.99\textwidth]{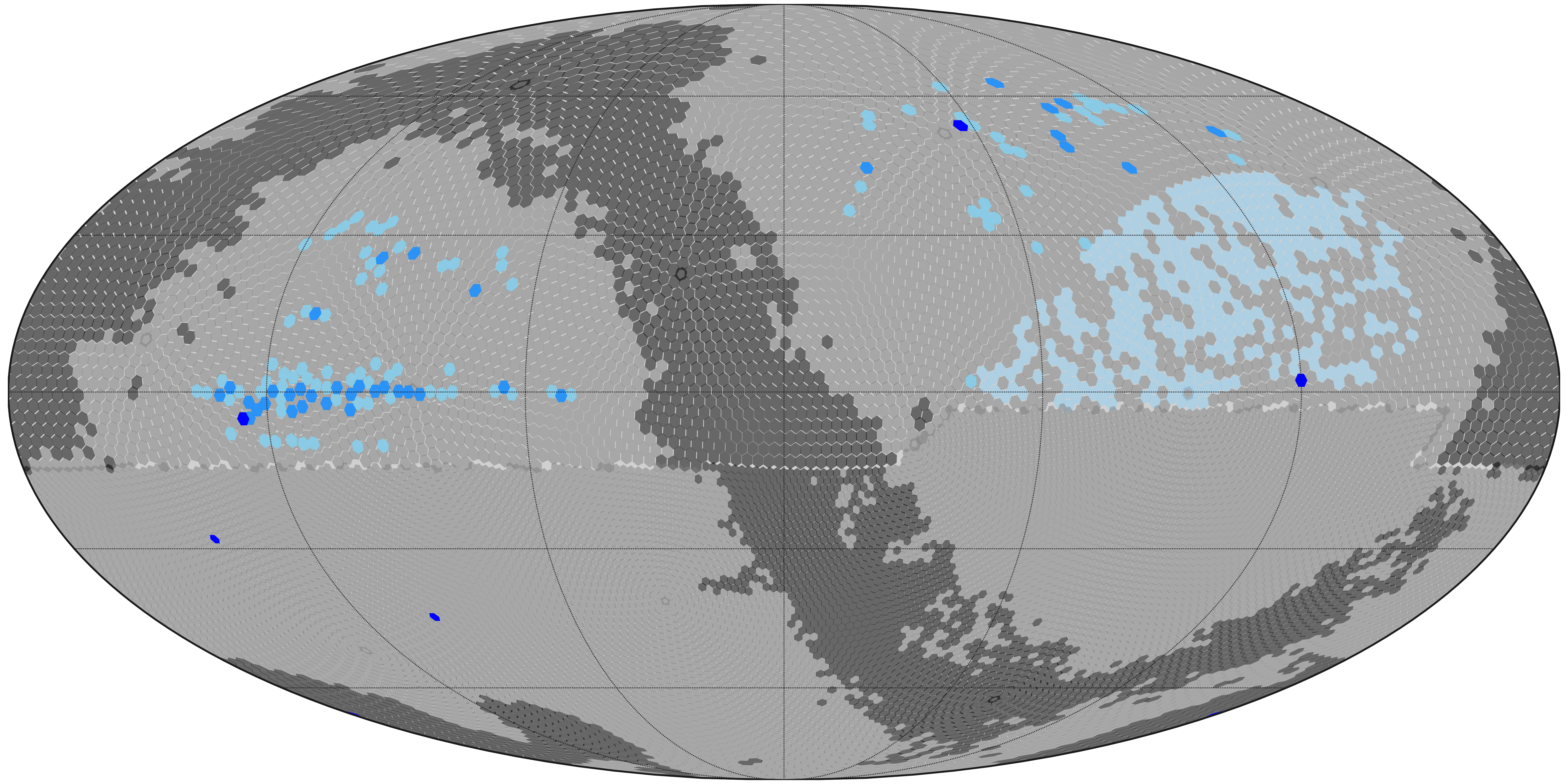}
\end{center}
\caption{\label{fig:fields} Field coverage in Equatorial coordinates,
  including the all-sky tiling plus specially defined fields. RA$=270$
  deg is at the center, and RA increases to the left. The difference
  in size between the LCO fields (in the South) and the APO fields (in
  the North) is apparent. Grey fields have no a priori constraints on
  their chosen cadence. Dark grey fields are forced to have
  observations; lighter grey fields are optional for {\tt
    robostrategy}, and we only plot those fields which actually are
  planned for observations. Blue fields have constraints on their
  cadence imposed by BHM; from lightest to darkest below they are
  AQMES Wide {\tt dark\_1x4} fields, AQMES Wide {\tt dark\_2x4}
  fields, AQMES Medium {\tt dark\_10x4} fields, and RM fields.}
\end{figure}

We use a set of fixed field positions on the sky, depicted in Figure
\ref{fig:fields}.  Most of the fields are in a tiling designed to
cover the whole sky.  A minimal covering set of fields is difficult to
find on a sphere; in two flat dimensions for our robotic positioner
coverage a hexagonal tiling would cover the area efficiently, but on a
2-sphere a fully hexagonal tiling (without gaps or overlaps between
fields) is not possible.  We use the spherical packing software due to
Hardin, Sloane, \& Smith, in particular their ``best covering'' sets
of locations.\footnote{\url{http://neilsloane.com/icosahedral.codes/}}
For the fields at APO, we take field locations as a subset of the
covering with 7,682 locations across the whole sky; for the fields at
LCO, we use a covering with 18,752 locations across the whole sky.

We divide the sky between APO and LCO in the following way:
\begin{itemize}
\item A field from the 7,682-location covering is used
at APO if its center satisfies:
\begin{itemize}
\item $\delta>-14^\circ$, and 
\item $0^\circ < l < 180^\circ$ or $b<10^\circ$ or $\delta > -3^\circ$.
\end{itemize}
\item A field from the 18,752-location covering is used
at LCO if its center satisfies:
\begin{itemize}
\item $\delta<-14^\circ$, or 
\item $180^\circ < l < 360^\circ$ and $b>10^\circ$ and $\delta < -3^\circ$.
\end{itemize}
\end{itemize}
This division splits the sky roughly into north and south sides, but
extends the LCO survey to the north in the Northern Galactic Cap. APO
is assigned a larger area because of the larger field of view of the
Sloan Foundation Telescope relative to the du Pont Telescope.

We run {\tt robostrategy} separately for each observatory with this
fixed division. It is possible in principle to use the {\tt
  robostrategy} methods to assign areas of sky to each observatory in
a way that maximizes performance. However, because of the differing
field size, doing so introduces extra complications that would have
taken more software development time and effort than available to us
to address.

For this all-sky tiling, we set the position angle of the field to
approximately match the surrounding hexagonal pattern.  The algorithm
we use is to search for the surrounding set of neighbors, with centers
less than twice the radius of the field of view, which typically
yields six neighbors (occasionally five). Then we set the position
angle to match the direction towards the furthest among those
neighbors. Because the position angles available at LCO only range
between 180$^\circ$ to 360$^\circ$, we restrict the allowed range of
position angles at LCO; we use the range from 240$^\circ$ to
300$^\circ$, which is sufficient given the six-fold near-symmetry of
the FPS system.

For the all-sky tiling fields, we have forced {\tt robostrategy} to
observe \newbf{at least a subset of them} with some cadence.
\newbf{For {\tt zeta-0} and {\tt iota-1}, every field was required to
  have a cadence. For {\tt theta-1}, we only guaranteed that fields were}
chosen to at least sparsely cover the entire range of Galactic
latitudes and longitudes, to cover the LMC and SMC, and to cover
important star forming regions.

In addition to these fields, a number of field locations are set
according to where desired targets are. The Reverberation Mapping
program uses the SDSS-RM field that has been observed continuously
since 2013 (\citealt{shen15a}), and three of Rubin Observatory's deep
drilling fields: the COSMOS field (\citealt{scoville07a}), the
XMM-Large Scale Structure field (\citealt{pierre07a}), and the Chandra
Deep Field South (\citealt{giacconi02a}). These fields all have
copious multiwavelength data and past and future photometric
monitoring.

A set of fields for the BHM AQMES Wide and Medium programs were chosen
in a process outside of robostrategy, via an optimization scheme
taking into account of information available during early survey
planning (in early 2020).  The number and placement of AQMES fields
was a compromise between maximizing the numbers of potential bright
($16<i_{\mathrm{psf}}<19.1$) SDSS DR16Q QSOs \citep{lyke20a} and
Chandra Source Catalog (CSC, version 2.1; \citealt{csc24a}) targets
per field, while also satisfying several other criteria, including the
expected demands on APO dark time from other SDSS-V programs. The
AQMES science requirement to follow up known SDSS QSOs limits us
naturally to the DR16Q footprint (i.e. Northern and Equatorial at high
Galactic latitude which is sky visible from APO).

The AQMES-Medium fields are few (36), and demand a relatively large
amount of exposure time per field. Therefore, to maximise the number
of targets per field, their centers were chosen from a fine grid of
potential sky positions.  Specifically, we chose the 36 highest
(non-overlapping) peaks (8 fields from the North Galactic Cap, 28 from
the SGC) in a smoothed HEALPix density map (NSIDE=512) of the number
of suitable QSOs and CSC targets (weighting these by a factor 12 and
0.8 respectively in creating the map). The balance between hemispheres
was chosen to better distribute the time request over a range of LSTs.

The AQMES-Wide fields are more numerous (hundreds) and have a natural
overlap with SPIDERS science.  Therefore, for the majority (330) we
limited their placement to field centers coinciding with the all-sky
tiling fields described above and within the $\sim$ 3000\,deg$^2$
overlap between the DR16Q sample and the ``SPIDERS'' Hemisphere
(approx. $180 < l < 360$, where the German eROSITA team have data
rights).  Fields were chosen which had the largest number of suitable
QSOs and CSC targets (here weighted 2 and 0.8 respectively). For these
fields we have planned starting in {\tt theta-1} for a single epoch
(cadence {\tt dark\_1x4}); in the original {\tt zeta-0} allocation
these fields were planned as {\tt dark\_2x4}. Finally, an additional
95 AQMES-wide fields outside the SPIDERS footprint were also
chosen. The centers of these fields were selected from peaks in a
smoothed HEALPix density map (NSIDE=512) of the number of suitable
QSOs and CSC targets (weighted 3 and 0.8 respectively), taking care to
avoid previously assigned AQMES fields.  For these fields we have
planned for two epochs (cadence {\tt dark\_2x4}) For each tier
described above, relative weighting between potential AQMES and CSC
targets was based on the relative exposure time request per target
that was expected at that time.

Figure \ref{fig:fields} shows the field positions and coverages. The
hexagons in the figure show the rough coverage of the robotic
positioners in each field. The radii of the approximate hexagon
vertices are 315 mm, corresponding to approximately 1.422 deg at APO
(covering 5.26 deg$^2$) and 0.952 deg at LCO (covering 2.35
deg$^2$). The quoted areas are approximations of the actual coverage
area; this area is slightly smaller for APOGEE than for BOSS.

% For SDSS-V we have used the Antiprism package to generate sets of
%field centers in approximately hexagonal patterns on a
%sphere.\footnote{\url{https://www.antiprism.com} }

\subsection{Total observing time availability}
\label{sec:observing_time}

We assume a particular time line for the project and evaluate every 15
minute increment to determine the LST and sky brightness
distributions. We use the {\tt PyAstronomy} software to determine the
Moon and Sun positions (\citealt{czesla19a}).  

The available night time is evening twilight to morning twilight. For 
``dark'' time twilight is when the Sun reaches 15 deg below the horizon.
For ``bright time'' the definition twilight depends on whether the 
Sun is above or below the Celestial Equator and on observatory. At APO,
bright twilight is at 8 deg when the Sun is above the Celestial Equator
($\sim$ March 21 to September 23) and 12 deg otherwise ($\sim$ September 
23 to March 21), and the reverse holds for LCO. This choice prevents 
overly long nights during the winter and is taken because of limitations 
on observer working hours.

We call ``dark'' the times when the Moon has a fractional illumination
less than 0.35 or is below the horizon, and when the Sun is more than
15 deg below the horizon. We call ``bright'' the times when the Moon
has a fractional illumination greater than 0.35 and is above the
horizon, and when the Sun is between 8 deg and 15 deg below the
horizon. Note that the survey scheduler at the telescope uses a
different, more detailed set of criteria accounting for Moon position
relative to the planned field and the resulting sky brightness.

The project time line assumes planned engineering time, which is
generally scheduled during full Moon, with an extended shutdown period
in the summer for APO.

Based on historical experience, we assume an effective fraction of
time that the telescope is open in nominally good conditions is 0.5
for APO and 0.7 for LCO. Actual open dome time tends to be on average
about 10\% larger than these assumptions; however, not all of that
time is in good conditions, and assuming these fractions for time in
nomimally good conditions is a good approximation to the experiences
of SDSS-III and SDSS-IV at APO and LCO.

\begin{figure}[!t]
\begin{center}
\includegraphics[width=0.49\textwidth]{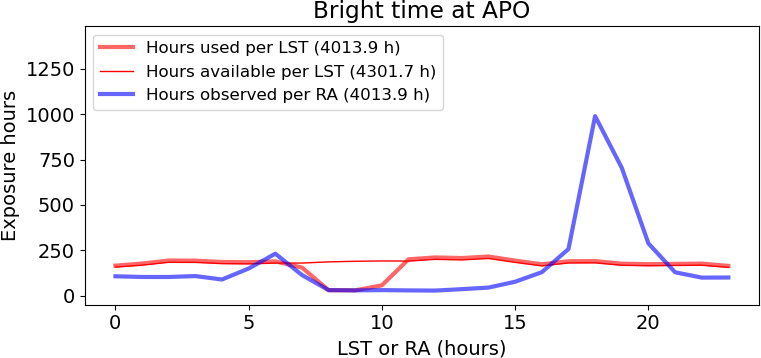}
\includegraphics[width=0.49\textwidth]{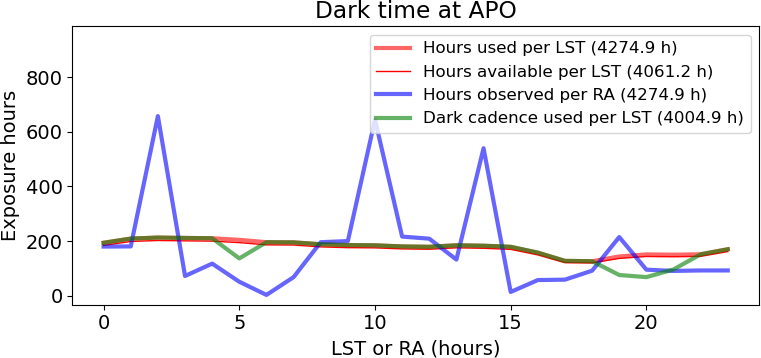}
\includegraphics[width=0.49\textwidth]{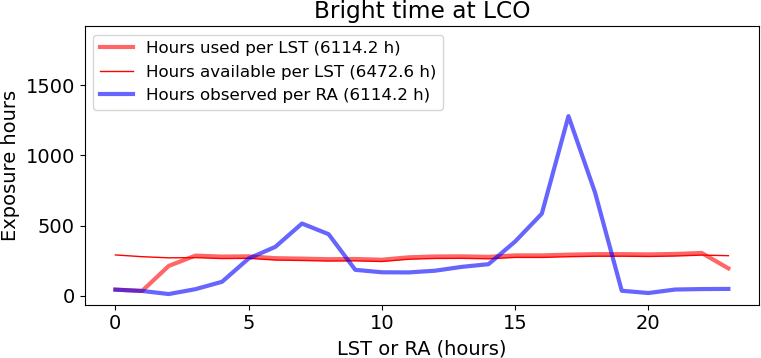}
\includegraphics[width=0.49\textwidth]{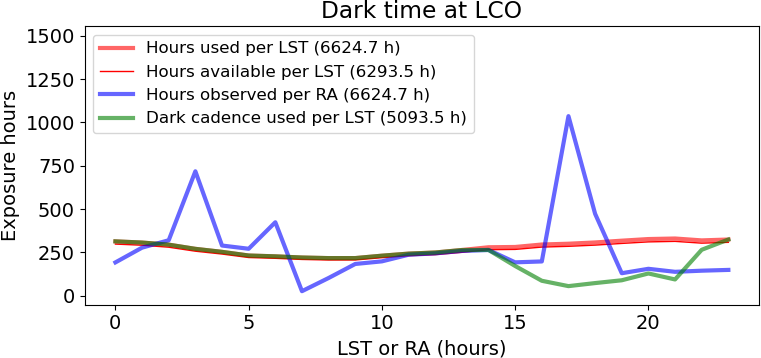}
\end{center}
\caption{\label{fig:time} Availability and allocation of time in {\tt
    zeta-0}. The left panels show bright time (as defined in the text)
  and the right panels show dark time. The top panels show APO and the
  bottom panels show LCO. The available time as a function of LST is
  the thin red line. The planned use of time as a function of LST
  based on the results is shown as the thick red line. The
  distribution of time spent at specific RAs are shown as the blue
  line. For dark time, the green line indicates the amount of time
  spent on dark cadence fields.  The Galactic Plane near 18h is
  apparent, particularly in bright time, but also in dark time at LCO,
  and the RM fields are apparent in dark time.}
\end{figure}

We divide LST into 24 hours, and using these definitions and
assumptions calculate the dark and bright time available for each
slot.  Figure \ref{fig:time} shows the available time as the thin red
line in each panel.

\subsection{Observing slot availability for each field}

For each field and cadence choice, we can observe it only during
certain of the slots shown in Figure \ref{fig:time}.

Bright time cadences can be observed in bright time or dark time, but
dark time cadences can only be observed in dark time.

The LST at which a field can be observed is subject to airmass
limits. In dark time there is an airmass limit of 1.4 in order
to minimize the effects of atmospheric diffraction, and in bright
time there is an airmass limit of 2 at APO and 1.7 at LCO. The limit
at LCO is determined by a lower altitude limit on the flat field
screen needed from BOSS calibrations, and may be subject to change if
the flat field screen infrastructure is adjusted in the future. If a
field center's declination has a minimum airmass greater than these
values, it is allowed to be observed, but can exceed its minimum
airmass only by 0.2.

The different limits for dark and bright are motivated by the fact
that the dark time cadences are generally designed for optical BOSS
observations whereas the bright time cadences are generally designed
for infrared APOGEE observations.  The necessary exposure time to
reach a given signal-to-noise ratio rises linearly with airmass for
BOSS observations, but based on an analysis of SDSS-III and SDSS-IV
data has no detectable dependence on airmass for APOGEE observations.

\begin{figure}[!t]
\begin{center}
\includegraphics[width=0.99\textwidth]{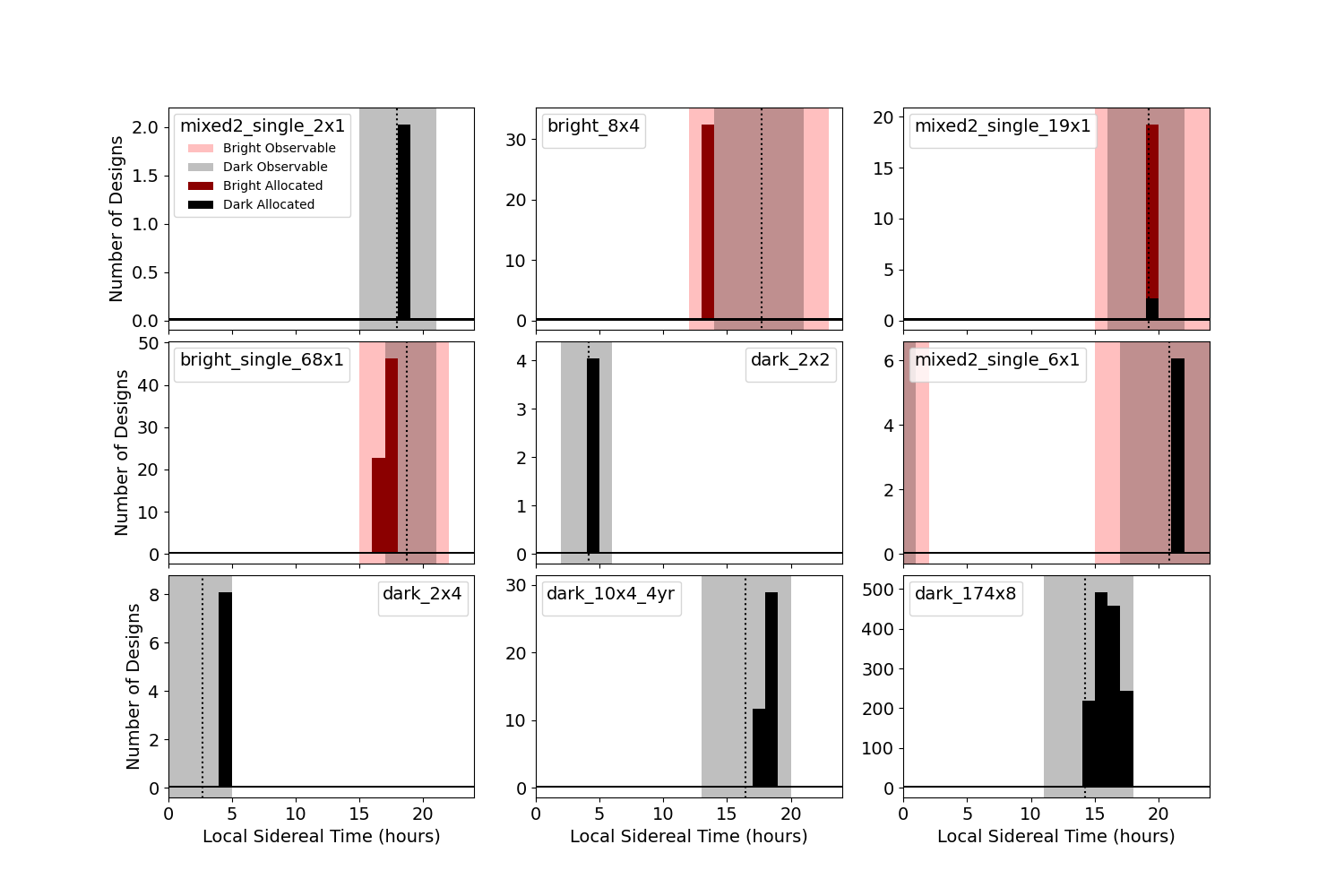}
\end{center}
\caption{\label{fig:timefields} Observability and allocation of time
  for nine individual fields and cadences (the cadences they were
  assigned). The {\tt mixed} cadences consist of two dark time designs
  followed by bright time designs. In each panel, the dotted vertical
  line corresponds to the RA of the field (i.e. the LST at which it
  transits). The pink and grey bands show the allowable LSTs in bright
  and dark time, which differ because of the different airmass limits.
  The dark time cadences only allow dark time, so only grey bands
  appear. The red and black histograms show the actual allocated
  distribution of LST resources in bright and dark time, respectively.
  In some cases dark time will be assigned even for bright time
  designs (e.g. for the four of the designs in the {\tt
    mixed2\_single\_6x1} cadence).
% from rs_paper_allocations
}
\end{figure}

Figure \ref{fig:timefields} shows the resulting available LSTs in
bright and dark time (as pink and grey bands, respectively) for nine
different fields and cadences.

\subsection{Observing efficiency and overhead assumptions}

Our time allocation needs to account for the relative efficiency of
observing as a function of airmass, and for the overheads involved in
observing. These factors translate into the factor $x_{ijk}$ in
Equation \ref{eq:basic} in Section \ref{sec:allocation}.  This factor
is calculated as:
\begin{equation}
\label{eq:cost}
x = \frac{1}{N_{\rm exp}} \left[t \times \left(\mathrm{airmass}^\alpha\right) + N_{\rm epochs} o_{\rm epoch} + 
N_{\rm exp} o_{\rm exp} \right]
\end{equation}
where $t$ is the time for a single observation (i.e. design), $N_{\rm
  exp}$ is the total number of designs in the cadence, $N_{\rm
  epochs}$ is the total number of epochs in the cadence, $o_{\rm
  epoch}$ is the overhead time associated with each epoch, and $o_{\rm
  exp}$ is the overhead time associcated with each observation.

High airmass tends to reduce efficiency because it degrades seeing,
reduces transparency, and increases chromatic differential
refraction. These effects are all more significant in the ultraviolet
and optical (i.e. for BOSS) than in the near infrared (i.e. for
APOGEE). High airmass also increases the sky emission foreground,
which more significantly affects targets faint relative to the sky
(which our optical targets tend to be) than targets bright relative to
the sky (which our infrared targets tend to be).  Estimates from past
BOSS observations suggest that to reach a given signal-to-noise
threshold requires an amount of observing time that scales linearly
proportional to airmass, i.e.  $\alpha=1$ in Equation \ref{eq:cost} 
(see also \citealt{dawson13a}).
Estimates from past APOGEE observations suggest that the necessary
observing time scales very weakly with airmass; we impose a dependence
proportional to airmass$^{0.05}$, i.e. $\alpha=0.05$ in Equation
\ref{eq:cost}, so as to prefer low airmass if all else is equal.

Overheads when observing must account for the slewing of the telescope
to a new field, the reconfiguration of the robots, and any
calibrations. Some of the overheads are only incurred once in an
epoch; e.g. during a visit to a particular field on a particular
night, the telescope only slews there once. Others are incurred each
design; e.g. if targets change between designs, the robots must be
reconfigured.

%The assumptions used for the {\tt theta-1} reallocation are
%conservatively defined. The net effect of our assumptions are that
%each dark design uses 24.9 minutes of open dome time at APO and 37.7
%minutes at LCO, and each bright design uses 22.3 minutes of open dome
%time at APO and 26.3 minutes of open dome time at LCO. 

\subsection{Calibration requirements}
\label{sec:calibration}

For the purposes of SDSS-V, a number of calibration fibers need to be
included: standard stars to estimate responses and sky locations to
estimate sky background, both in APOGEE and BOSS fibers.  {\tt
  robostrategy} receives these targets as special cartons generated by
the target selection software.  The calibration requirements are not
accounted for in the field cadence allocation; they are only used
during the final fiber assignment step.

\citet{medan25a} describes the motivation behind and the specific
choices for the calibration requirements. These choices include the numbers of
sky and standard targets as well as the magnitude limits of standards under
different design modes.  Additionally, as described in Section
\ref{sec:greedysrd}, because of their relatively small number, the
APOGEE standard stars have a required focal plane distribution.
We also have a preferred distribution of APOGEE standard colors and
magnitudes (implemented in {\tt zeta-3} and later), and fully
described in \citet{medan25a}.

%The standards are sorted by their goodness, where:
%\begin{equation}
%%\mathrm{goodness} = - 13.33\left[(J-K) - 0.25\right] - (H - 9),
%\end{equation}
%which prefers brighter and bluer stars. This condition was set based
%on experiments comparing how well telluric features determined for
%individual stars in a field agreed with the smooth spatial fit
%determined from all stars in the field, as a function of brightness
%and color of the individual stars. The residuals measured in these
%experiments showed a trend in brightness and color roughly
%corresponding to the above equation. At the beginning of the fiber
%assignment for any field, we determine the goodness threshold to use
%for selecting APOGEE standard in each zone in Figure \ref{fig:fps}. We
%use a goodness threshold of zero by default, but if there are less
%than 3 stars available above the threshold, we reduce the threshold
%until there are (or no more standards are available). The assignment
%process (which requires assigning at least one APOGEE standard to each
%zone, and 15 APOGEE standards total) then has enough to choose from
%among the best available APOGEE standards.

%\input{bright-limits}
%
%\input{calibrations}

\subsection{Bright neighbor conditions}
\label{sec:brightneighbor}

Bright stars in the field can contribute light to fibers accidentally,
which can cause saturation or excess scattered light in the
instruments. To reduce this effect, we ensure that no fiber on an
assigned robot positioner is placed near a bright
star. \citet{medan25a} describes the details of these exclusion areas
and the observational tests performed to validate them.  We check both
the BOSS and (if it is connected to the spectrograph) the APOGEE
fiber, and prevent an assignment if the relevant bright limit is
violated in either case.

Our method does not prevent star or galaxies from contributing
substantially to the flux down the fiber; after all, we only exclude
locations for which that flux would exceed the bright
limit. Furthermore, we do not account for nearby extended galaxies at
all. Finally, we do not check unassigned robot positioners, which are
a small fraction of all robot positioners, particularly in parts of the 
sky that have a higher surface density of potentially contaminating 
bright stars..

\subsection{Offsets}
\label{sec:offsets}

To accommodate very bright targets in certain cartons, we implement
fiber offsets, which prevent saturation and undue scattered light in
the spectrographs. These offsets allow us to probe much brighter stars
than previously possible with the BOSS and APOGEE spectrographs using
the 2.5-m telescopes, though the spectrophotometry on these stars is
unreliable. \citet{medan25a} describe in detail how these offset
amounts are determined. They essentially offset the fiber by the
amount necessary to satisfy the bright neighbor conditions, with an
extra safety factor.

% We use the same functions for offsets that we used for the bright
% neighbor test, but with an extra safety factor applied of 0.5 mag in
% bright time and 1 mag in dark time. Here, the safety factor is added 
% to the magnitude difference needed to get the bright source at the 
% magnitude limit for the design. Therefore the offset always
% exceeds the safe distance from the bright target.

% Figure \ref{fig:offset} shows the offset as a function of magnitude
% for BOSS and APOGEE at both observatories.  For APOGEE, the function
% is the same in all design modes. For BOSS, the bright limits differ,
% so we show separately the {\tt bright\_time}, {\tt dark\_plane}, and
% the other dark time design modes. The offsets are finite even just
% under the bright limit, to ensure safety from saturation and scattered
% light.

% The relevant magnitude for APOGEE is $H$. We exclude entirely any
% object with $H<1$. For BOSS, we check all of the available magnitudes
% for which a bright limit is defined (see Table
% \ref{table:bright-limits}) and use the largest calculated offset. We
% exclude any object for which any of these magnitudes are $<6$ in
% bright time, and $<13$ in dark time.

We apply the offset in the positive right ascension direction, which is roughly
perpendicular to the parallactic direction for observations taken at
transit. This choice reduces (but only partially) the effects of
the offset on spectrophotometry, while retaining operational
simplicity.

% {\bf Ilija really needs to check.}

% \begin{figure}[!t]
% \begin{center}
% \includegraphics[width=0.95\textwidth]{offset_definition.png}
% \end{center}
% \caption{\label{fig:offset} Fiber offsets as a function of target
  % magnitude. For BOSS, the offset applied is the maximum calculated
  % from the $g$, $r$, $i$, $B_P$, $G$, or $R_P$ magnitudes; for APOGEE,
  % the offset is applied as a function of the $H$ magnitude. Each
  % different line represents the offset for a given design mode at a
  % given observatory, as labeled. The offsets take effect brighter than
  % the bright limits in each mode, as listed in Table
  % \ref{table:bright-limits}. There is an absolute limit to the
  % brightness, which is the magnitude at which each line terminates. At
  % small offsets, the offset applied at LCO is smaller than at APO
  % because of the typically better atmospheric seeing, which dominates
  % the core of the PSF.}
% \end{figure}

\subsection{Assignment stages}
\label{sec:assignment_stages}

The fiber assignment methods described in Section \ref{sec:assignment} are applied in
SDSS-V in several stages, as alluded to earlier. These stages reflect
the relative priorities of observing different groups of targets:
\begin{itemize}
\item SRD: assignment of the targets that drive the science requirements
of the survey.
\item Reassignment: assigning extra observations to already-assigned
SRD targets or partial cadences to unassigned SRD targets.
\item Open Fiber: assigning open fiber targets.
\item Filler: assigning filler targets.
\item Complete: assignments to fill any unassigned fibers.
\end{itemize}

\subsubsection{SRD Stage}

The SRD stage assigns the SRD stage targets (see Section
\ref{sec:inandout}) and the calibration targets, using the method
described in Section \ref{sec:greedysrd}. At this point, the SRD
targets are fixed. The calibration targets can exceed their
requirements at this stage, and may be removed in later stages if they
can be used on science targets. 

\subsubsection{Reassignment Stage}

The reassignment stage is designed to increase the scientific output
of the survey by assigning fiber resources to SRD targets with relaxed
cadence requirements and a different observing philosophy than the
previous stage. At the SRD stage, each carton assigns a single observing
cadence to a target, selected to achieve a particular science goal, and {\tt
  robostrategy} either made assignements to fully satisfiy the cadence
or made no assignments. However, for some cartons, a subset of the
science goals may still be achievable if less than the full cadence
specification of epochs and exposures are completed. For other
cartons, the cadence reflects the minimum observations needed to
achieve the science, but additional observations would enhance the
science goals by providing either improved signal-to-noise or
additional time-sampling. Thus, reassignment cartons are those for
which ``partial completion'' and/or ``extra completion'' has high
scientific value. Partial and extra completion can either be at the
epoch level (in which case the number of exposures per epoch defined
by the SRD cadence must be available for assignment to occur) or the
expoure level (in which case any available exposure in any epoch could
be assigned). No constraints on epoch separations are considered, but
lunation requirements are respected. In the {\tt zeta} series of {\tt
  robostrategy} runs, there was no limit to the number of extra
exposures or epochs that could be assigned at this stage. In later
{\tt robostrategy} runs, each carton set a cap so that the number of
total exposures or epochs was $\sim2 \times$ the number specified by
the target cadence.  Cartons identified for reassignment are worked
through sequentially, in an order agreed upon by the mappers, but
within this ordering target priorites are respected.  Science cartons
not identified for the reassignment stage may still have opportunities
for leftover fiber resources in the last stage of {\tt robostrategy}.

The cartons that have an opportunity to obtain extra epochs of 
observation for completed targets at this stage are:
\begin{itemize}
 \item {\tt bhm\_gua\_dark}
 \item {\tt bhm\_gua\_bright}
 \item {\tt mwm\_astar\_core\_boss}
 \item {\tt mwm\_cb\_cvcandidates\_boss}
 \item {\tt mwm\_cb\_galex\_vol\_boss}
 \item {\tt mwm\_cb\_galex\_mag\_boss}
 \item {\tt mwm\_ob\_core\_boss}
 \item {\tt mwm\_snc\_100pc\_boss}
 \item {\tt mwm\_snc\_ext\_main\_boss}
 \item {\tt mwm\_tess\_2min\_apogee}
 \item {\tt mwm\_wd\_pwd\_boss}
 \item Any carton in the {\tt bhm\_csc} program
 \item Any carton in the {\tt bhm\_spiders} program
 \item Any carton in the {\tt mwm\_magcloud} program
 \item Any radial velocity-related carton (identified as having ``{\tt \_rv\_}'' in its name)
\end{itemize}

The cartons that are eligible for partial completion (if they
could not be originally completed) at this  stage are:
\begin{itemize}
  \item {\tt bhm\_gua\_bright}
  \item {\tt bhm\_gua\_dark}
  \item {\tt manual\_mwm\_validation\_rv\_apogee}
  \item {\tt mwm\_astar\_core\_boss}
  \item {\tt mwm\_bin\_rv\_long\_apogee}
  \item {\tt mwm\_ob\_core\_boss}
  \item Any carton in the {\tt bhm\_csc} program
  \item Any carton in the {\tt bhm\_spiders} program
  \item Any carton in the {\tt mwm\_yso} program
\end{itemize}

\subsubsection{Open Fiber Stage}

The open fiber target stage assigns individual exposures to a set of
target categories that were submitted by collaboration members as
being of special interest (see Section 8 of \citealt{almeida23a}). 
They are limited to single-observation
cadences but can choose dark or bright sky conditions. They are assigned 
using the greedy method described in
Section \ref{sec:greedy}.

The filler target stage is similar to the open fiber target stage, and
effectively these targets are just lower priority open fiber
targets. However, the filler cartons by design are very large sets of
targets covering as much of the sky as possible, in order to fill
spare fibers with as many potentially useful science targets as possible.

The ``complete'' stage is the final stage, and tries to fill any
remaining unassigned fibers. First, we find cases where science
targets are assigned for partial epochs, i.e. for only some of the
designs in a given epoch. For all such cases, in order of priority, we
try to assign the science targets to the remaining designs in those
epochs. Second, we then find any science targets that are assigned in
at least one design, and in order of priority try to assign them in
any other designs (across epochs). Third, we do the same for any unassigned science targets,
which gives the opportunity to observe partial cadences. Fourth, we
then do the same for calibration targets. We apply the method to the
standard star categories first, and then to the sky targets, which
means that any unassigned fibers at the very end have the opportunity
to be put on a known sky location (rather than a random part of sky,
which would be the fate of a completely unassigned fiber).

In four dedicated reverberation mapping (RM) fields, we perform the 
target assignments through a separate process, using the constraint programming 
assignment algorithm described in  Section \ref{sec:constraintassignment} to 
maximize the number of observed quasars at each priority level. 
Only BOSS calibration 
fibers are necessary in RM fields because no APOGEE fibers are assigned. 
For some epochs, targets other than quasars (for example, white dwarf 
stars) are assigned fibers to  pursue non-RM science.

\subsection{Results}
\label{sec:results}

Here we present some results for the {\tt robostrategy} runs up
to the middle of SDSS-V.

Figure \ref{fig:time} shows the allocated distribution of dark and
bright time at each observatory (as estimated at the beginning of 
SDSS-V FPS operations) as a function of LST, as the thin red
line. The RA distribution of the observations is shown as the blue
line. The RM fields and the Galactic Plane cause considerable
non-uniformity in the RA distribution, and and matching this
demand to the available LST resources
leads to fields being observed at higher airmass (i.e. while rising or 
setting instead of at transit). The green line shows the dark time which is allocated
to dark cadences, demonstrating that observing the Galactic Plane
requires a substantial amount of the dark time at those LSTs.

\begin{figure}[!t]
\begin{center}
\includegraphics[width=0.49\textwidth]{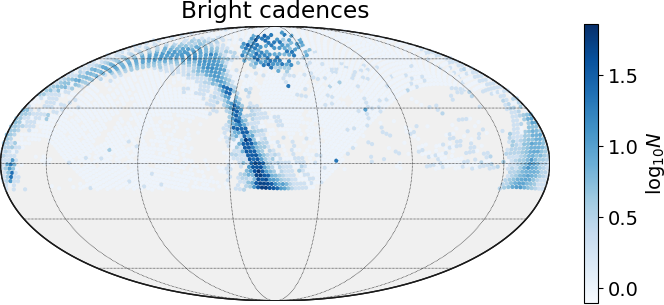}
\includegraphics[width=0.49\textwidth]{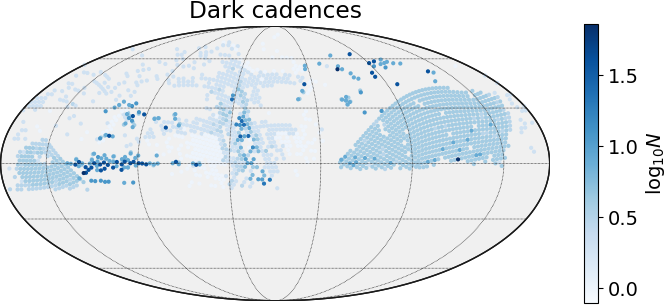}
\includegraphics[width=0.49\textwidth]{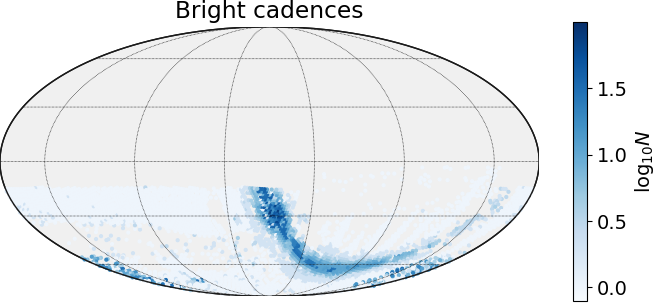}
\includegraphics[width=0.49\textwidth]{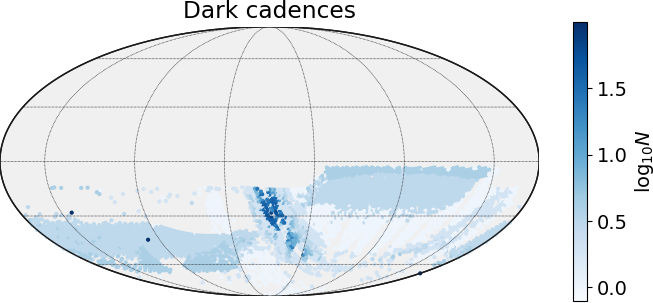}
\end{center}
\caption{\label{fig:allocation-sky} Distribution of number of planned
  observations $N$ on the sky in dark and bright time at each
  observatory (top is APO, bottom is LCO). The sky projection is the
  same as Figure \ref{fig:targets}. These are the allocations from
  the {\tt iota-1} run of {\tt robostrategy}.}
\end{figure}

Figure \ref{fig:allocation-sky} shows the allocation of dark and
bright time across the sky for each observatory, under the {\tt
  iota-1} plan. This plot shows that the planned sky distribution of observations is very
heterogenous; note particularly the many observations conducted in the
Galactic Plane fields, driven by the large target density there. 

\begin{figure}[!t]
\begin{center}
\includegraphics[width=0.98\textwidth]{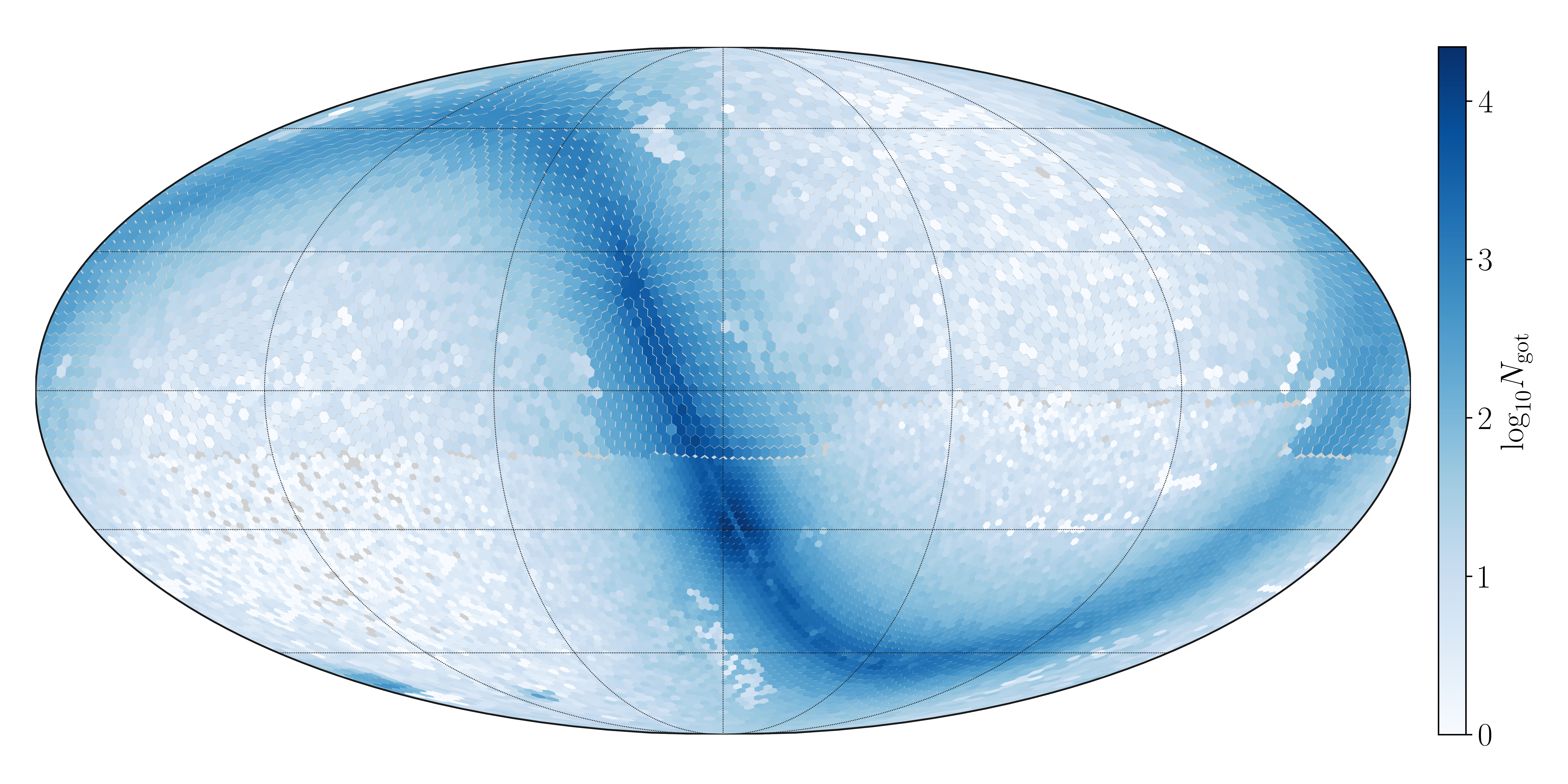}
\includegraphics[width=0.98\textwidth]{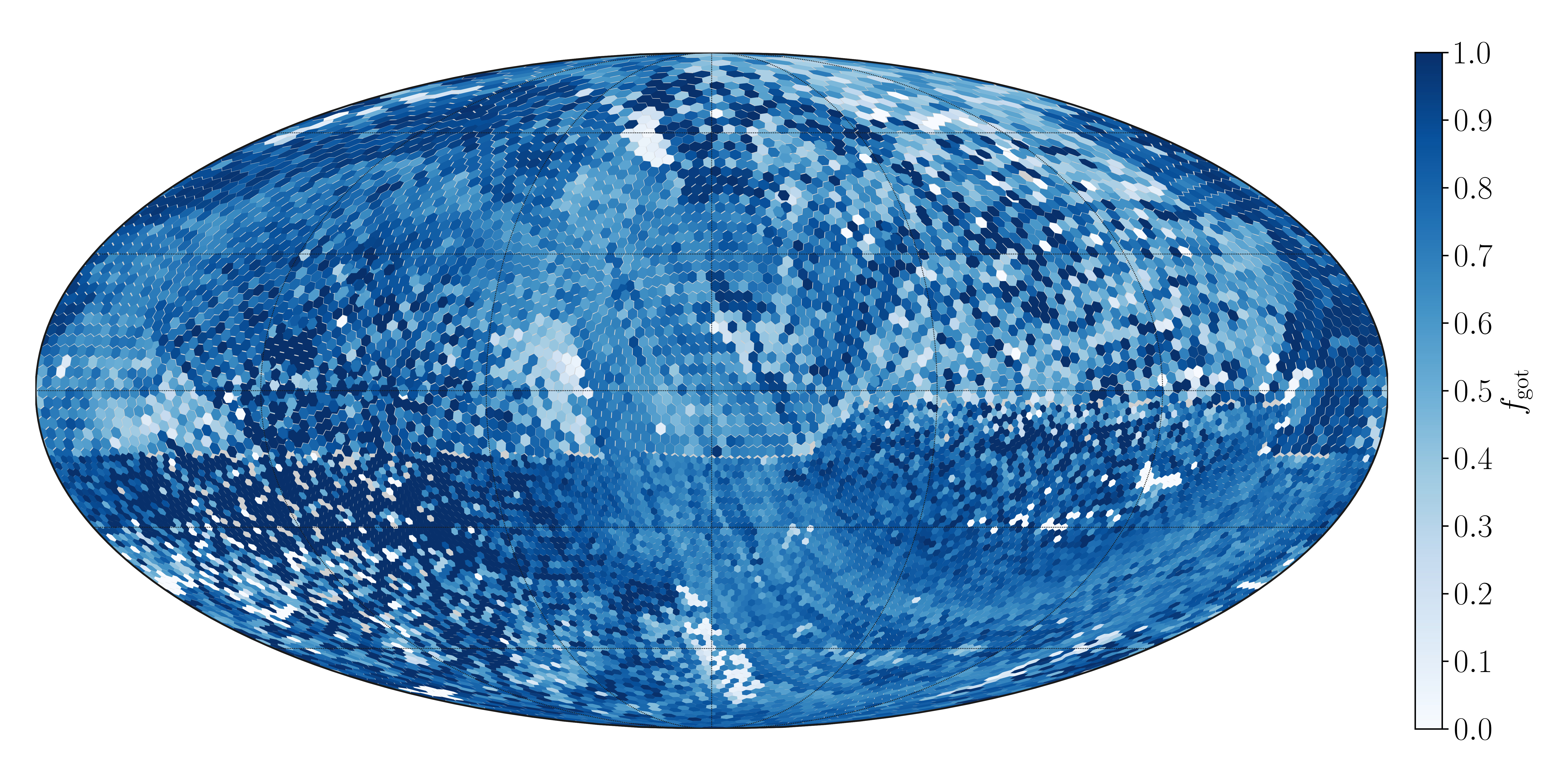}
\end{center}
\caption{\label{fig:ggcompleteness} For each field, number of assigned
  targets (top) and fraction of assigned targets relative to the total
  number of targets (bottom) for the {\tt
    mwm\_galactic\_core\_dist\_apogee\_sparse} carton in run {\tt
    iota-1}. The sky projection is the same as Figure
  \ref{fig:targets}.  Grey areas are those with no field coverage. The
  areas of greatest incompleteness are associated with areas observed
  early in the survey that used a different target selection; most of
  the current targets were not identified at that time.}
\end{figure}

Figure \ref{fig:ggcompleteness} shows how this plan assigns 
targets to the {\tt mwm\_galactic\_core\_apogee\_sparse} carton, the
largest SRD carton in terms of numbers of targets. The completeness is
reasonably, though far from completely, uniform across the sky. Some
dips in completeness at mid-Galactic latitudes is notable, and is
associated with fields with a small number of total designs; in these
situations, adding designs is relatively expensive and inefficient.

\begin{figure}[!t]
\begin{center}
\includegraphics[width=0.48\textwidth]{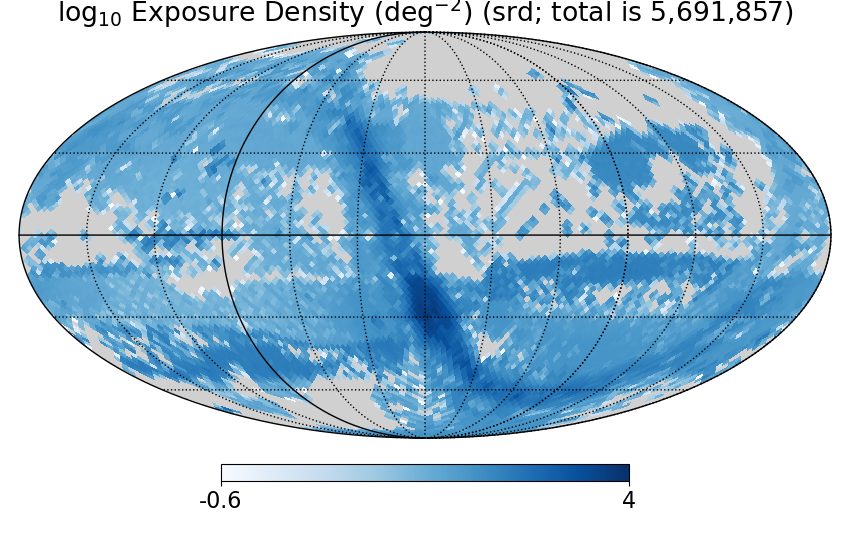}
\includegraphics[width=0.48\textwidth]{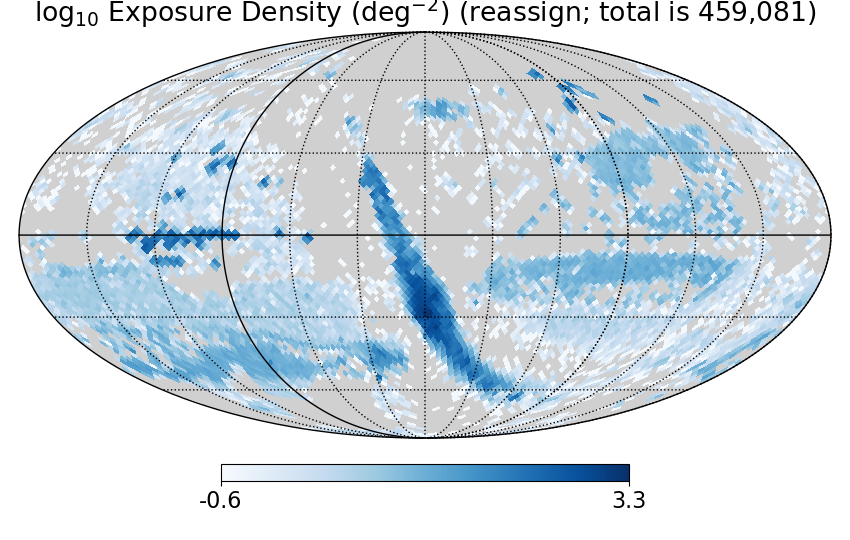}
\includegraphics[width=0.48\textwidth]{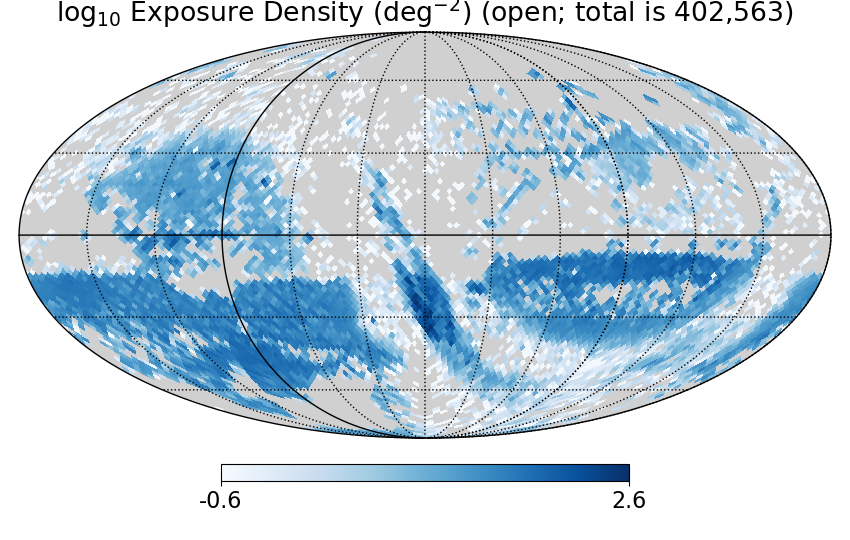}
\includegraphics[width=0.48\textwidth]{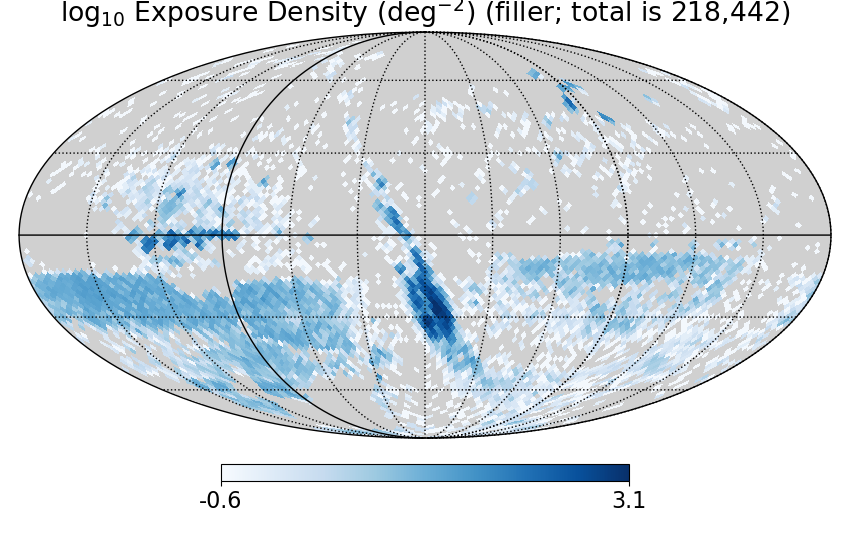}
\includegraphics[width=0.48\textwidth]{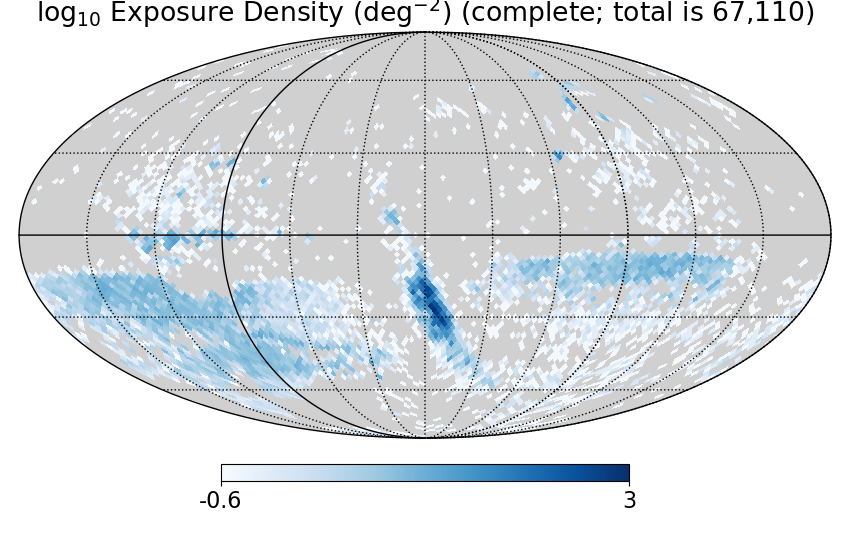}
\includegraphics[width=0.48\textwidth]{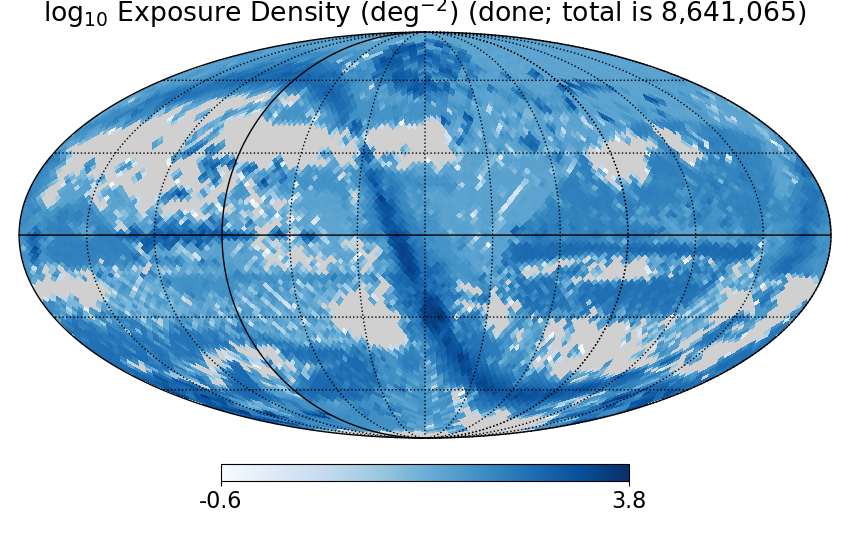}
\end{center}
\caption{\label{fig:stages} Sky distribution of fiber-exposures newly assigned at
  time of the {\tt iota-1} reassignment. Top left
  panel shows the exposures assigned during the SRD step. Top right
  panel shows the exposures assigned during the reassignment
  step. Middle left panel shows the exposures assigned during the open
  fiber target process. Middle right panel shows the exposures
  assigned during the filler target process. Bottom left panel shows the
  exposures assigned during the ``complete'' process. Bottom right
  panel shows the exposures already taken by the time of {\tt
    iota-1}. The sky
  projection is the same as Figure \ref{fig:targets}. }
\end{figure}

Figure \ref{fig:stages} shows how the exposures are assigned across
each of the fiber assignment stages, during the replanning {\tt
  robostrategy} performed in {\tt iota-1}. At the time of this
reassignment, many exposures had already been taken, and are depicted
in the bottom right. During the reassignment, the majority of
exposures are in the SRD stage, but many other exposures are assigned
in later stages as well. The open, filler, and complete maps indicate
how much spare capacity there is in the system.

\begin{figure}[!t]
\begin{center}
\includegraphics[width=0.48\textwidth]{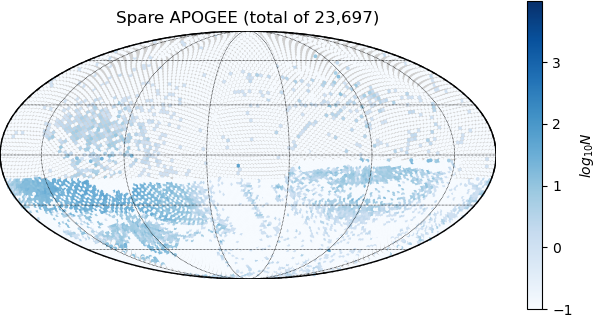}
\includegraphics[width=0.48\textwidth]{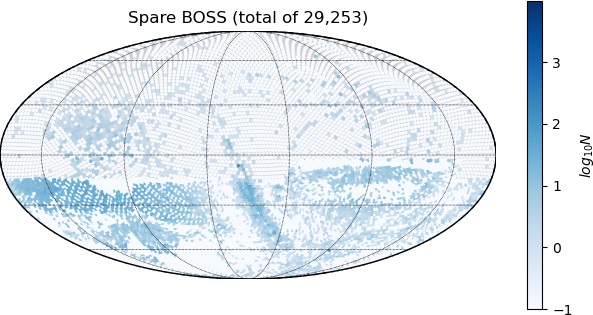}
\end{center}
\caption{\label{fig:spares} Sky distribution of spare APOGEE fibers
  (left) and BOSS fibers (right), excluding Reverberation Mapping
  fields. The majority of these fibers (about 60\%) were able to be
  assigned as calibrations. The set of BOSS spare fiber positioners is
  a superset of the set of APOGEE spare fiber positioners. The sky
  projection is the same as Figure \ref{fig:targets}. }
\end{figure}

Figure \ref{fig:spares} shows the distribution of spare fiber
observations for the APOGEE and BOSS instruments. Because every fiber
positioner with a APOGEE fiber also has a BOSS fiber, the count of
spare BOSS fibers includes all the spare APOGEE fibers. These maps
exclude the Reverberation Mapping fields, which have only a small
number of spare fibers. The spare fiber count includes fibers that are
assigned to ``spare calibration targets'' (as described in Section
\ref{sec:greedysrd}), which constitute about 60\% of the spare
fibers. The number of spare or completely unused fiber exposures is
exceedingly small.

\section{Summary and Discussion}
\label{sec:summary}

We have described {\tt robostrategy}, the software used by SDSS-V to plan
observations for its FPS systems conducting multiobject spectroscopy
for the MWM and BHM programs. {\tt robostrategy} allocates cadences to
each of the survey fields in accordance with survey requirements and
available telescope time. Within each field, {\tt robostrategy}
assigns fibers to targets for each design, in such a way that if the
field is observed in its planned cadence, the cadences of individual
targets will be achieved as well. We have described the general
methods, which may prove useful for future surveys, and some of the
particulars associated with SDSS-V.

We expect only minor further development in {\tt robostrategy} over
the course of SDSS-V, though we do expect to perform more
reallocations to account for performance changes and weather
stochasticity over the course of the survey.

There are important lessons to draw from writing and optimizing this
code, regarding considerations for future programs. \newbf{A general
  point is that although our procedure for field allocation is
  near-optimal relative to our chosen metric, its use does not obviate
  the need to choose the values that different targets contribute to
  the metric, to explore the consequences of those choices, and to
  decide on scientific trade-offs through a detailed discussion. Thus,
  although an automated procedure is desirable given the scale of the
  survey, it still involves human judgment to set its input
  parameters.}

\newbf{There are other more specific lessons.} First, although we have
a system for dealing with nearly-arbitrary cadence requirement
specifications, doing so adds considerable complexity to the code. In
contrast, APOGEE and APOGEE-2 in SDSS-III and SDSS-IV implemented a
much more simplified set of cadence possibilities. The trade-off
between complexity and handling arbitary cadences is important to
consider, particularly in light of the upcoming deluge of time-domain
targets at all cadences. Second, and similarly, the large number of
target cartons in SDSS-V, while part of the core scientific mission of
the program, have added considerably to preparation time to each
update of the survey plan. Third, the {\tt robostrategy} approach of
planning the entire survey simultaneously causes a bottleneck in
implementing changes in target selection and strategy\newbf{, because
  a full run takes several days from start to finish}. It is possible
that a more hierarchically-designed method would allow more rapid,
modular changes to subsets of the program. Finally, there are some
designs with {\tt bright\_time} observing modes that we end up
observing in dark time. In principle, there are targets we could
include on those designs that require dark time; however, implementing
this optimization would be a bit complex to avoid changing the set of
highest priority bright cadence targets that are assigned on these
designs.

For future surveys, {\tt robostrategy} provides a flexible method for time
allocation and fiber assignment that incorporates many different
targeting categories and accounts for available observing time as a
function of LST and lunation. There are improvements and extensions
that would be useful to consider. For fiber assignment, applying a
constraint programming approach (Section
\ref{sec:constraintassignment}) to all fields would improve fiber
assignment efficiency and could lead to greater algorithmic
simplicity; but doing so requires greater computational power or more
efficient solvers. Accounting for targets in overlapping fields could
lead to greater efficiency (\citealt{blanton03a}).  Optimizing field
placement could also lead to improvements (\citealt{blanton03a,
  tempel20b}); doing so in fact requires accounting for targets in
overlapping fields. Finally, as noted above, optimizing APO and LCO
time allocations together could lead to a much better division of sky
area between the two observatories. 

\newbf{The {\tt robostrategy} software used for this implementation is
  available on
  GitHub.}\footnote{\url{https://github.com/sdss/robostrategy}}

\begin{acknowledgements}

\newbf{We thank Lynne Jones for useful feedback at the
beginning of planning for this work, as well as
Keith Inight, Alexandre Roman Lopes, and an anonymous referee
for 
useful comments on the manuscript.}

Funding for the Sloan Digital Sky Survey V has been provided by the
Alfred P. Sloan Foundation, the Heising-Simons Foundation, the
National Science Foundation, and the Participating Institutions. SDSS
acknowledges support and resources from the Center for
High-Performance Computing at the University of Utah. SDSS telescopes
are located at Apache Point Observatory, funded by the Astrophysical
Research Consortium and operated by New Mexico State University, and
at Las Campanas Observatory, operated by the Carnegie Institution for
Science. The SDSS web site is \url{www.sdss.org}.

SDSS is managed by the Astrophysical Research Consortium for the
Participating Institutions of the SDSS Collaboration, including the
Carnegie Institution for Science, Chilean National Time Allocation
Committee (CNTAC) ratified researchers, Caltech, the Gotham
Participation Group, Harvard University, Heidelberg University, The
Flatiron Institute, The Johns Hopkins University, L'Ecole
polytechnique f\'{e}d\'{e}rale de Lausanne (EPFL), Leibniz-Institut
f\"{u}r Astrophysik Potsdam (AIP), Max-Planck-Institut f\"{u}r
Astronomie (MPIA Heidelberg), Max-Planck-Institut f\"{u}r
Extraterrestrische Physik (MPE), Nanjing University, National
Astronomical Observatories of China (NAOC), New Mexico State
University, The Ohio State University, Pennsylvania State University,
Smithsonian Astrophysical Observatory, Space Telescope Science
Institute (STScI), the Stellar Astrophysics Participation Group,
Universidad Nacional Aut\'{o}noma de M\'{e}xico, University of
Arizona, University of Colorado Boulder, University of Illinois at
Urbana-Champaign, University of Toronto, University of Utah,
University of Virginia, Yale University, and Yunnan University.

This work is based on data from eROSITA, the soft X-ray instrument
aboard SRG, a joint Russian-German science mission supported by the
Russian Space Agency (Roskosmos), in the interests of the Russian
Academy of Sciences represented by its Space Research Institue (IKI),
and the Deutsches Zentrum für Luft- und Raumfahrt (DLR). The SRG
spacecraft was built by Lavochkin Association (NPOL) and its
subcontractors, and is operated by NPOL with support from the
Max-Planck Institute for Extraterrestrial Physics (MPE).

The development and construction of the eROSITA X-ray instrument was
led by MPE, with contributions from the Dr.\ Karl Remeis Observatory
Bamberg, the University of Hamburg Observatory, the Leibniz Institute
for Astrophysics Potsdam (AIP), and the Institute for Astronomy and
Astrophysics of the University of T\"ubingen, with the support of DLR
and the Max Planck Society. The Argelander Institute for Astronomy of
the University of Bonn and the Ludwig Maximilians Universit\"at Munich
also participated in the science preparation for eROSITA.

\software{Astropy \citep{astropy13a},
  Matplotlib \citep{hunter07a},
  NumPy \citep{harris2020array},
  SciPy \citep{2020SciPy-NMeth},
  Google OR-Tools \citep{ortools},
  PyAstronomy \citep{czesla19a}}

\end{acknowledgements}

\bibliography{blanton}{}
\bibliographystyle{aasjournal}



\end{document}